\documentclass[a4paper]{aa}
\usepackage{graphicx}
\usepackage{txfonts}
\usepackage{color}
\usepackage{natbib}
\usepackage{lineno}
\bibpunct{(}{)}{;}{a}{}{,} 

\def\aap{Astron. Astrophys.}

\def\ao{Applied Optics}

\def\apjl{Astrophys. J. Letters}

\def\apjs{Astrophys. J. Suppl.}

\begin{document}


\title{Inferring asymmetric limb cloudiness on exoplanets from transit light curves}
\titlerunning{Clouds and asymmetric transits}

\author{P. von Paris\inst{1,2}   \and P. Gratier\inst{1,2} \and  P. Bord\'e\inst{1,2}  \and J. Leconte\inst{1,2} \and F. Selsis\inst{1,2} }

\institute{Univ. Bordeaux, LAB, UMR 5804, F-33270, Floirac, France \and CNRS, LAB, UMR 5804, F-33270, Floirac, France 
}

\abstract {Clouds have been shown to be present in many exoplanetary atmospheres. Cloud formation modeling predicts considerable inhomogeneities of cloud cover, consistent with optical phase curve observations. However, optical phase curves cannot resolve some existing degeneracies between cloud location and cloud optical properties.}{We present a conceptually simple technique to detect inhomogeneous cloud cover on exoplanets. Such an inhomogeneous cloud cover produces an asymmetric primary transit of the planet in front of the host star. Asymmetric transits produce characteristic residuals compared to a standard symmetric model. Furthermore, bisector spans can be used to determine asymmetries in the transit light curve.}{We apply a model of asymmetric transits to the light curves of HAT-P-7b, Kepler-7b and HD209458b and search for possible cloud signatures. The nearly uninterrupted Kepler photometry is particularly well-suited for this method since it allows for a very high time resolution.}{We do not find any statistically sound cloud signature in the data of the considered planets. For HAT-P-7b, a tentative detection of an asymmetric cloud cover is found, consistent with analysis of the optical phase curve. Based on Bayesian probability arguments, a symmetric model with an offset in the transit ephemeris remains, however, the most viable model. Still, this work demonstrates that for suitable targets, namely low-gravity planets around bright stars, the method can be used to constrain cloud cover characteristics and is thus a helpful additional tool to study exoplanetary atmospheres.}{}


\maketitle

\section{Introduction}

Many observations reported in recent years are compatible with the presence of clouds in exoplanetary atmospheres. 

Transmission spectra measure the transit depth during primary transit as a function of wavelength. If the atmosphere is opaque at a given wavelength, the planet will appear larger at that wavelength. However, visible, near-IR and IR transmission spectra of planets such as GJ1214b (e.g., \citealp{bean2010}, \citealp{kreidberg2014}), GJ436b (e.g., \citealp{knutson2014}), CoRoT-1b (e.g., \citealp{schlawin2014}), WASP-31b (e.g., \citealp{sing2015}) or HAT-P-32b (e.g., \citealp{gibson2013}), show very little variation with wavelength. They can be approximated by flat lines to within measurement uncertainties. This has been interpreted as being due to optically thick cloud layers high up in the atmosphere, which mask the expected molecular and atomic absorption.

During secondary eclipse, the apparent dayside brightness temperature can be measured at IR wavelengths. Again, as different wavelengths probe different atmospheric pressure due to wavelength-dependent opacity, brightness temperatures are expected to vary. As was observed for transmission spectra, IR emission spectra of many exoplanets are consistent with being featureless (e.g., \citealp{hansen2014}). A single blackbody brightness temperature can be assigned to the photosphere, independent of wavelength. Again, this is consistent with uniform cloud coverage that blocks the access to deeper atmospheric layers.

UV-visible secondary eclipse spectra have been used to determine the wavelength-dependent albedo of hot Jupiters (e.g., \citealp{evans2013}). It is found that the slope in the albedo spectrum is consistent with clouds obscuring deeper atmosphere levels where molecular and atomic absorption could influence the observations.

The CoRoT and Kepler space missions have provided a wealth of high-precision, broadband optical photometry of thousands of exoplanets and candidates. About 20 of these exoplanets show detectable phase curves (e.g., \citealp{snellen2009}, \citealp{mazeh2010}, \citealp{barclay2012}, \citealp{quintana2013}, \citealp{esteves2013,esteves2015}), i.e. variations with orbital phase as the planetary dayside (which reflects/re-emits starlight back to the observer) rotates in and out of view. A few Kepler phase curves show asymmetries with respect to secondary eclipse. Post-eclipse maxima of the phase curves have been found for at least four exoplanets (e.g., \citealp{demory2013inhomogen}, \citealp{esteves2015}, \citealp{webber2015}). These are conceptually interpreted to be a consequence of inhomogeneous cloud cover on the planet dayside. Simply put, clouds form on the nightside and evaporate as they are transported from the cooler "morning" side across the substellar meridian to the hotter "evening" side\footnote{Morning and evening refer here to an assumed global eastward atmospheric circulation.}.

In recent years, detailed theoretical cloud modeling has been performed (e.g., \citealp{sudarsky2000}, \citealp{parmentier2013}, \citealp{wakeford2015}, \citealp{webber2015}, \citealp{lee2015}). These studies used temperature structures calculated by 1D and 3D atmospheric models to predict cloud condensate composition (e.g., silicate and iron clouds have been proposed) and cloud altitude (usually around 1-100\,mbar of pressure). Together with 3D circulation patterns, sophisticated cloud formation schemes simulate sedimentation and cloud particle growth to calculate cloud particle size distributions, cloud particle densities and cloud location. These studies generally confirm the heterogeneous cloud cover inferred from optical phase curves. However, given current data quality, degeneracies exist between cloud parameters that are not easily remedied with optical phase curves alone. For instance, even homogeneous cloud cover (or clouds clustered towards the substellar point) could produce asymmetric phase curves if the scattering asymmetry factor of the cloud particles is adapted accordingly.

Therefore, apart from intensive numerical modeling efforts, additional diagnostics for cloud properties are needed to help resolve the degeneracies and better constrain cloud characteristics. In this work, we propose a simple concept to probe cloud locations, i.e., asymmetric primary transits\footnote{While submitting this manuscript for publication, we became aware of another paper that also advocates this idea \citep{line2015}. However, they focus on the transmission spectra and do not concentrate further on photometric data.}. The concept is based on the premise that, in the case of heterogeneous cloud cover, the western limb of the planet (the leading limb in primary transit) is cloudy, whereas the eastern limb (trailing limb) is cloud-free, as proposed based on phase curve observations. Thus, the planet is composed of two hemispheres with different apparent radii, which produces different transit signatures during ingress and egress.

Transit observations, especially from space-based surveys, are particular useful for this method for several reasons. Mainly, the quasi-continuous coverage allows for the analysis of hundreds of near-consecutive transits, hence an excellent time resolution of the order of seconds or less in the phase-folded light curves. Furthermore, compared to spectroscopic observations, broadband photometry produces much higher signal-to-noise ratios and, thus enables stronger constraints on cloud parameters.

We apply the new technique to two planets in the Kepler field with well-characterized asymmetric phase curves, namely Kepler-7b (e.g., \citealp{latham2010}, \citealp{demory2011kepler7,demory2013inhomogen}, \citealp{esteves2015}) and HAT-P-7b (\citealp{pal2008}, \citealp{borucki2009}, \citealp{welsh2010}, \citealp{esteves2015}, \citealp{vparis2015phase}). Furthermore, we analyze transit light curves of HD209458b taken from \citet{brown2001} (their Figure 3) obtained from four transits observed with the Hubble Space Telescope. HD209458 is one of the brightest stars around which a transiting planet has been found, and the transmission spectrum of HD209458b shows some evidence of possible cloud and/or haze cover (e.g., \citealp{deming2013}). Therefore, we include it here for illustration purposes.

Although the results are not conclusive with respect to the presence of clouds, we will establish the applicability of the technique for future photometric surveys.

The paper is organized as follows: Section \ref{datared} presents the data reduction process adapted for this work, and Sect. \ref{formodel} presents the numerical model used in the transit modeling. Section \ref{clouddiag} introduces the cloud diagnostics for the primary transit light curve. The inverse model used to fit the forward models to the data is described in Sect. \ref{invmodel}. Results are presented in Sect. \ref{results} and discussed in Sect. \ref{discuss}. We conclude with Sect. \ref{summary}.

\section{Data reduction}

\label{datared}

\begin{figure}[h]
\begin{center}
\includegraphics[width=250pt]{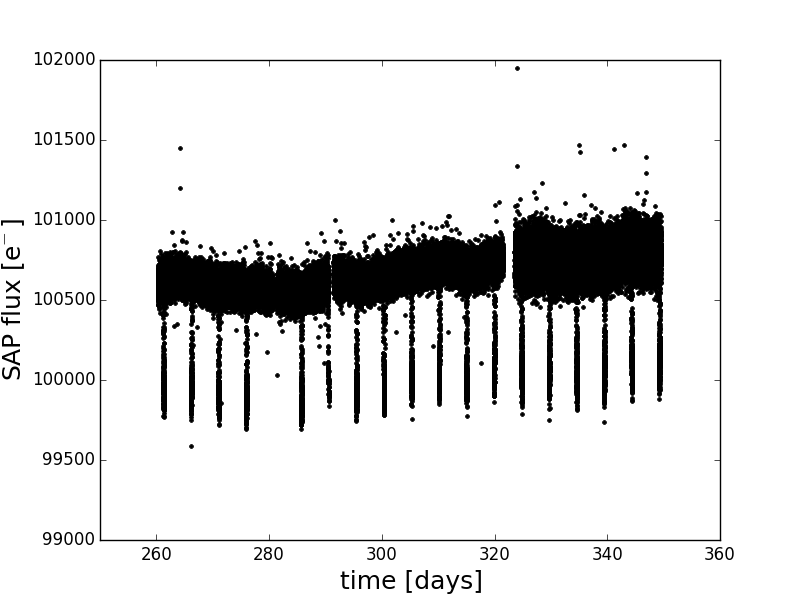}\\
\caption{SC-SAP for Kepler-7 in quarter Q3.}
\label{sap_q3}
\end{center}
\end{figure}

For target planets observed by Kepler (i.e., Kepler-7b and HAT-P-7b), we use the short-cadence (SC) single-aperture photometry (SAP), which is publicly available from the MAST archive\footnote{http://archive.stsci.edu/kepler/data\_search/search.php}. An example is shown in Fig. \ref{sap_q3}.

First, data points, that are flagged by the Kepler pipeline as NaNs and obviously flawed data, are removed manually to produce a raw light curve $R_{qk}$ for each quarter $k$. To obtain phase-folded transit light curves $L_{qk}$, we then proceed in three steps:

\begin{enumerate}

\item Outlier removal: We calculate a running median $M_{2.4}$ for each point $j$ of the light curve, using a 2.4-hour window. 

\begin{equation}
\label{median1hour}
M_{2.4}(j)=\rm{median}\left(R(i), |t(i)-t(j)|\leqslant 1.2 \rm{hrs}\right),
\end{equation}

where $t$ denotes the time stamps of the data point.

From there, we calculate the median absolute deviation (MAD) as follows:

\begin{equation}
\label{mad_def}
\rm{MAD}=\rm{median}\left(|R-M_{2.4}|\right).
\end{equation}

A point $j$ is rejected if

\begin{equation}
\label{reject}
|R(j)-M_{2.4}(j)|\geqslant 4\times \rm{MAD}.
\end{equation} 

Figure \ref{transit9} illustrates this procedure, for the 9th transit of Kepler-7b in Q3 (green lines).

\begin{figure}[h]
\begin{center}
\includegraphics[width=250pt]{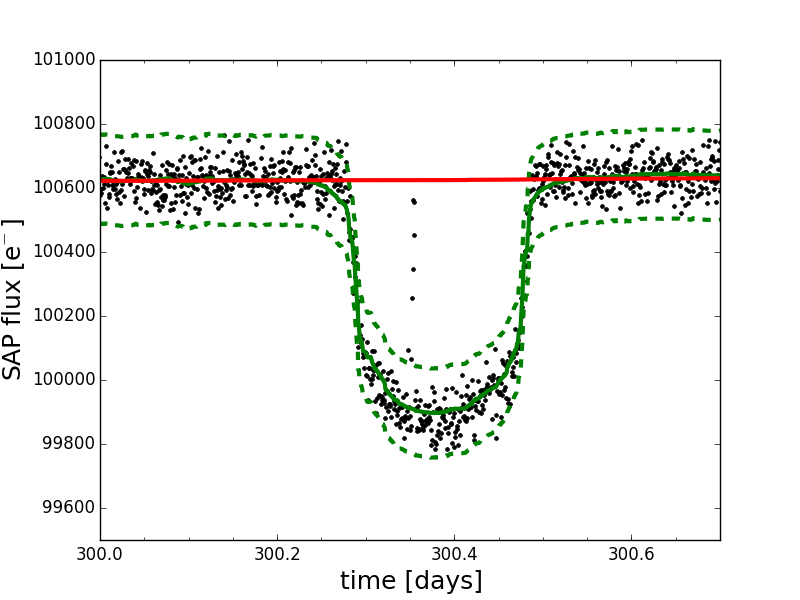}\\
\caption{9th transit in Q3: Illustration of running 2-day median (red line) and 2.4-hour median (green plain line). Green dashed lines encompass the 4x$\rm{MAD}$ criterion to identify outliers (see text).}
\label{transit9}
\end{center}
\end{figure}

\item Normalization: After outlier removal, we calculate a running median $M_{48}$ for each point $j$, using a roughly 2-day window (red line in Fig. \ref{transit9}).

\begin{equation}
\label{median1day}
M_{48}(j)=\rm{median}\left(R(i), |t(i)-t(j)|\leqslant 24 \rm{hrs}\right).
\end{equation}

The normalized light curve $N$ is obtained by dividing the raw light curve by the median-filtered light curve.

\begin{equation}
\label{normalized}
N=\frac{R}{M_{48}}.
\end{equation}

In this way, slow variations (due to, e.g., planetary phase curve, stellar activity, etc.) are mostly removed from the final light curve. Figure \ref{normalq3} shows the normalized light curve for Kepler-7 in quarter 3.

\begin{figure}[h]
\begin{center}
\includegraphics[width=250pt]{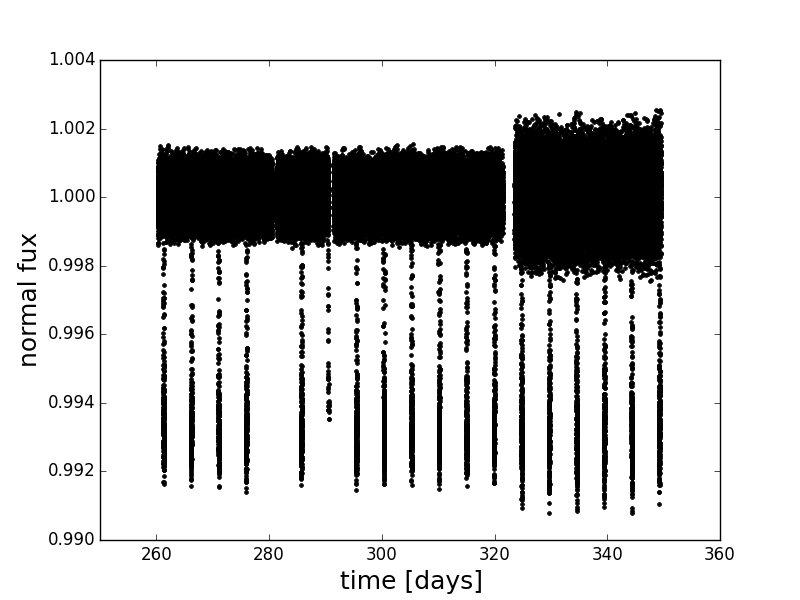}\\
\caption{Normalized SC-SAP for Kepler-7 in Q3.}
\label{normalq3}
\end{center}
\end{figure}

\item Phase folding: The normalized light curve $N$ is then phase-folded to obtain the final light curve $L$ as a function of orbital phase $\Phi$. For this, $\Phi$ is defined such that $\Phi$=0 at primary transit. 

\begin{equation}
\label{phasedef}
\Phi(j)=\frac{t(j)-T_0}{P}-\lfloor \frac{t(j)-T_0}{P} \rfloor,
\end{equation}

where $T_0$ is the mid-transit time of the first transit, $P$ the orbital period and $\lfloor x \rfloor$ represents the floor function, i.e., the  greatest integer less than or equal to $x$. Figure \ref{phasefoldedq3} shows the final light curve of Kepler-7b, using all quarters with SC data.

\end{enumerate}

\begin{figure}[h]
\begin{center}
\includegraphics[width=250pt]{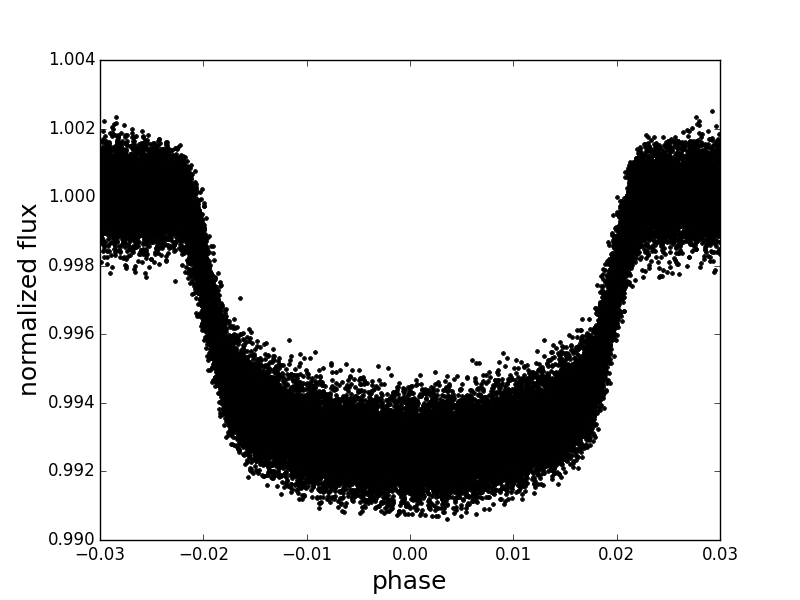}\\
\caption{Phase-folded SC-SAP for Kepler-7 in Q3-Q8.}
\label{phasefoldedq3}
\end{center}
\end{figure}

The values for $T_0$ and $P$ are taken from the latest data release (DR24)\footnote{http://exoplanetarchive.ipac.caltech.edu} and summarized in Table \ref{dr24}.

\begin{table}[h]
  \centering 
  \caption{DR24 values for $T_0$ and $P$, used in eq. \ref{phasedef}.}\label{dr24}
  \begin{tabular}{l|ll}
\hline
\hline
   Planet& $T_0$-2454833 [days] & $P$ [days] \\
\hline
  Kepler-7b &134.2768383 &4.885488953\\
   HAT-P-7b & 121.3585723 & 2.204735365
\end{tabular}
\end{table}

\begin{figure}[h]
\begin{center}
\includegraphics[width=250pt]{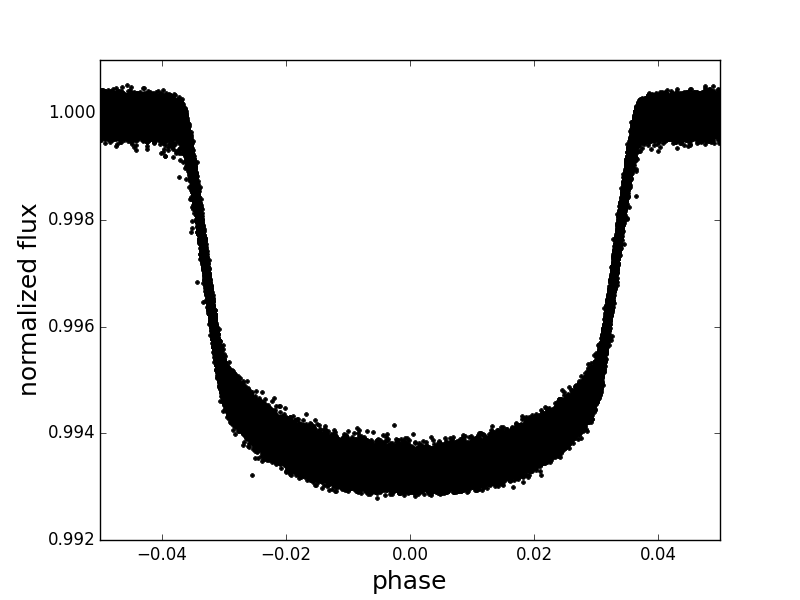}\\
\caption{Phase-folded SC-SAP for HAT-P-7b in Q0-Q17.}
\label{phasefoldedq3_hat}
\end{center}
\end{figure}

Figure \ref{phasefoldedq3_hat} shows the Q0-Q17 light curve of HAT-P-7b. Since HAT-P-7 is about 2.5 magnitudes brighter than Kepler-7 in the Kepler bandpass, the resulting light curve is much cleaner.

\begin{figure}[h]
\begin{center}
\includegraphics[width=250pt]{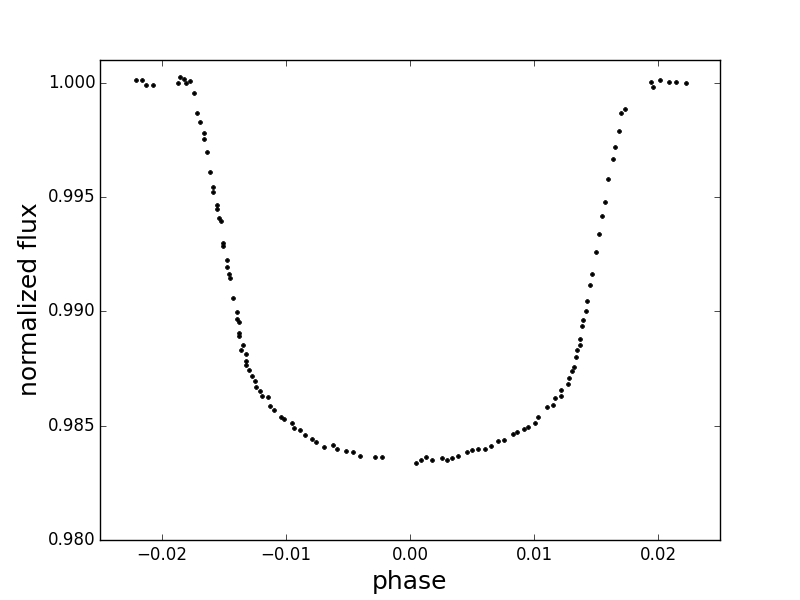}\\
\caption{Transit light curve of HD209458b. Data from \citet{brown2001}.}
\label{trans_hd}
\end{center}
\end{figure}

Figure \ref{trans_hd} shows the transit light curve of HD209458b, used in this work. Data is taken from \citet{brown2001} (their Figure 3). Observations were made with the Hubble Space Telescope during four transits of HD209458b and then phase-folded to obtain a composite light curve. As is clearly seen, the photometric quality of the light curve is exceptional, since HD209458 is a $V$=8$^m$ star and was observed with the HST (aperture 2.4\,m). However, an important point is the relatively low time resolution compared to the Kepler light curves.

\section{Forward model}

\label{formodel}

\label{toymodel}

\begin{figure}[h]
\begin{center}
\includegraphics[width=250pt]{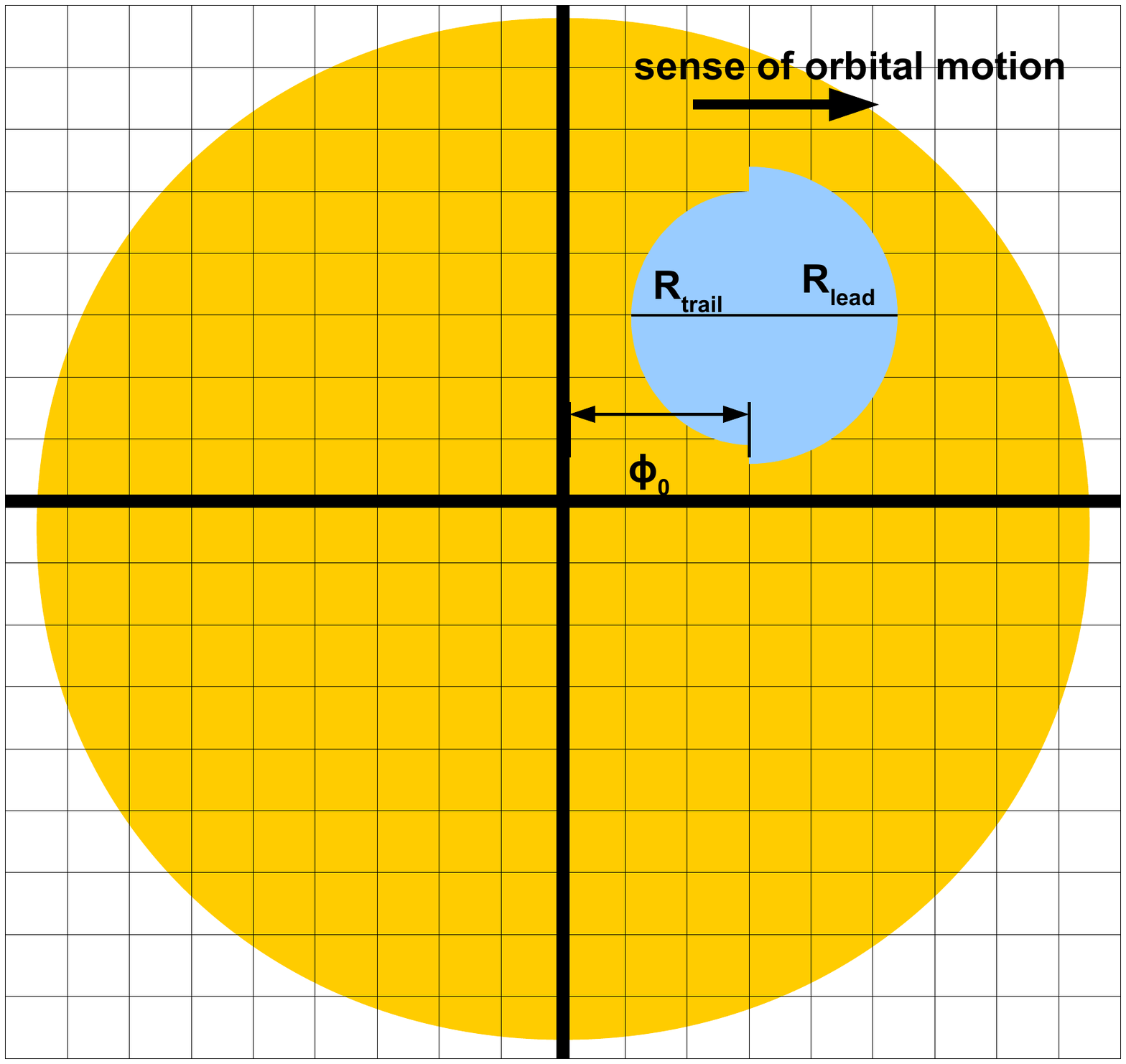}\\
\caption{Illustration of an asymmetric transit: The leading, cloudy hemisphere has a larger radius than the trailing, cloud-free hemisphere.}
\label{asymsketch}
\end{center}
\end{figure}

In this model, the planet is assumed to be composed of two hemispheres with radius $R_{\rm{trail}}$ and $R_{\rm{lead}}$ for the trailing and leading hemisphere, respectively. These two radii are parameterized as follows:

\begin{eqnarray}\label{hemispheres}
 R_{\rm{trail}}& = & R_p \\
R_{\rm{lead}} & = & R_p+ h_{\rm{cloud}}
\end{eqnarray}

where $R_p$ is the planetary radius and $h_{\rm{cloud}}$ is the cloud altitude. Note that $h_{\rm{cloud}}$<0 is permitted, which then implies a larger trailing hemisphere. The concept of the model is illustrated in Fig. \ref{asymsketch}.

To calculate the stellar total luminosity and estimate the area obscured by the planet, we use a $N_{\ast}\times N_{\ast}$ grid to resolve the stellar disk, with its center corresponding to the stellar center (see Fig. \ref{asymsketch}). $N_{\ast}$ is chosen such that a stellar pixel corresponds to roughly half a scale height of the planet in question ($N_{\ast}$=4,000 for Kepler-7b, $N_{\ast}$=10,000 for HAT-P-7b, for example).

As is general practice in exoplanet transit modeling, the asymmetric model adopts a quadratic limb-darkening law with $u_1$ and $u_2$ as the linear and the quadratic limb-darkening coefficients for the intensity distribution $I$ across the stellar disk:

\begin{equation}
\label{limbquad}
I(\mu)=1-u_1\cdot (1-\mu)- u_2\cdot (1-\mu)^2,
\end{equation}

with $\mu=\cos (\phi)$ and $\phi$ is the angle between the local surface normal and the observer direction.

Planetary orbits are assumed to be circular, which is a good approximation for the targets considered (see, e.g., \citealp{winn2009}, \citealp{demory2011kepler7,demory2013inhomogen}, \citealp{esteves2015}). 

Further parameters needed to describe the passage of the planet in front of the star are the impact parameter $b$ and $a_S$,  the projected semi-major axis $a_S$ in units of the stellar radius:

\begin{equation}
\label{kdef}
a_S=\frac{a}{R_{\ast}},
\end{equation}

where $a$ is the semi-major axis and  $R_{\ast}$ is the stellar radius (fixed in this work due to strong asteroseismology constraints, e.g. \citealp{demory2011kepler7}, \citealp{eylen2012}).

The impact parameter is defined as the projected distance (in units of stellar radii) of the planets' center to the stellar center at mid-transit. It is related to the orbital inclination $i$ via:

\begin{equation}
\label{impact_para}
b=\cos \left(i\right) \cdot a_S .
\end{equation}

In addition, we introduce a seventh parameter which is the phase offset $\phi_0$ of the passage of the planetary meridian over the stellar meridian (see Fig. \ref{asymsketch}) compared to the minimum of the primary transit. In general, we assume $\phi_0$=0, however due to, e.g., eccentric orbits, $\phi_0$ can be different from zero. 

The passage of the planet in front of the star is discretized such that the planet advances less than one stellar pixel per time step.

We have compared the new model with the widely used, analytical \citet{mandel2002} model. Usually, the deviations between both models are of the order of 1 ppm or less, hence much smaller than the signals we look for (see next section, Fig. \ref{residu_illu}). We deem such a deviation acceptable. With increasing numerical resolution in our model, the deviations will also decrease, however at a much increased computational cost.

In total, the asymmetric model contains up to seven free parameters, namely $R_p$, $h_{\rm{cloud}}$ to describe the form of the planet, $a_S$, $b$, $\phi_0$ to describe the orbit of the planet, and $u_1$, $u_2$ to describe the stellar intensity distribution. 

%
%

\section{Cloud diagnostics}

\label{clouddiag}

\subsection{Bisector span}

\begin{figure}[h]
\begin{center}
\includegraphics[width=250pt]{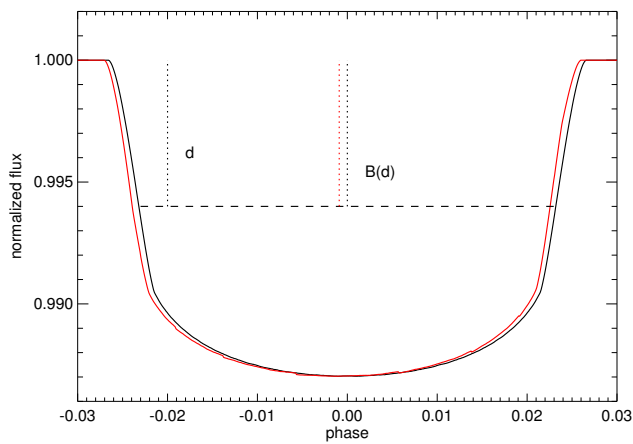}\\
\caption{Illustration of the definition of the bisector $B$ as a function of depth $d$. Symmetric transit in black, asymmetric transit with "morning" clouds in red.}
\label{bisect_def}
\end{center}
\end{figure}

The bisector span is the first potential diagnostic of an asymmetric transit, that could indicate the presence of clouds. The bisector $B$ as a function of transit depth $d$ is a geometrical line. It is defined as the center of a horizontal line that intersects the transit light curve at a given $d$ (see Fig. \ref{bisect_def} for an illustration).

For a symmetric transit of a uniform planet, the bisector is a straight, vertical line, centered at zero. However, for asymmetric planets, two phenomena are observed. Firstly, the bisector is non-zero, shifted by a few minutes compared to the cloud-free case. Secondly, it is no longer a straight line, rather, with transit depth, a triangular-like shape can be seen. This shape is caused by the varying slope of the light curve as cloudy and cloud-free hemispheres, respectively, begin/end their passage in front of the star.

\begin{figure}[h]
\begin{center}
\includegraphics[width=250pt]{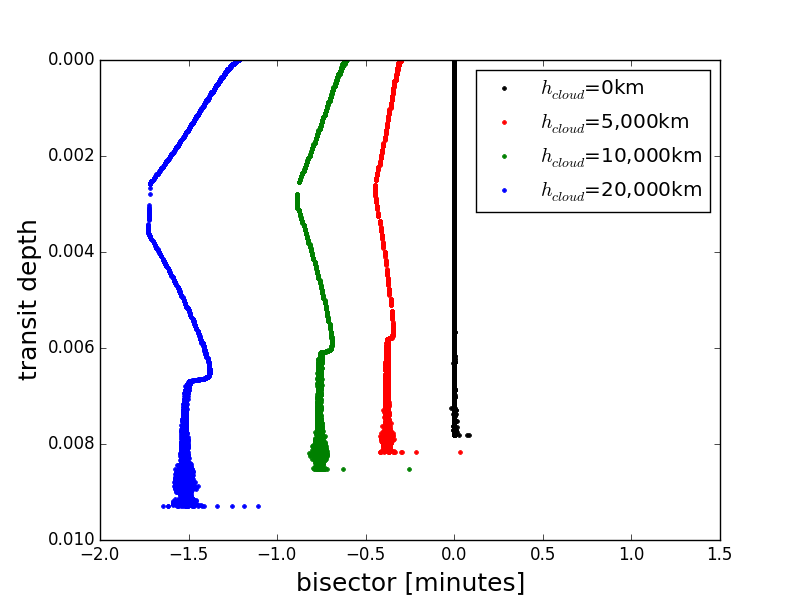}\\
\caption{Bisector values for different adopted $h_{\rm{cloud}}$. No observational noise present.}
\label{bisect}
\end{center}
\end{figure}

Figure \ref{bisect} illustrates the bisector span for different values of cloud altitudes (zero cloud altitude corresponding to a symmetric transit), as calculated by the asymmetric model (Sect. \ref{toymodel}). The adopted planetary and stellar parameters were those for Kepler-7b. Note the noise-like features at high transit depth, where the transit light curve is very flat ($d\gtrsim$0.008 for the blue line). These features are due to the finite numerical resolution in the model.

\begin{figure}[h]
\begin{center}
\includegraphics[width=250pt]{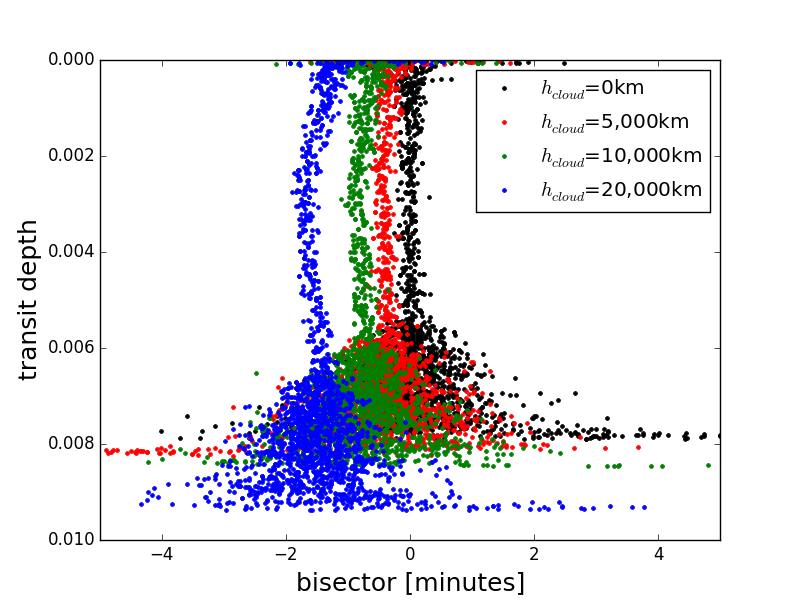}\\
\includegraphics[width=250pt]{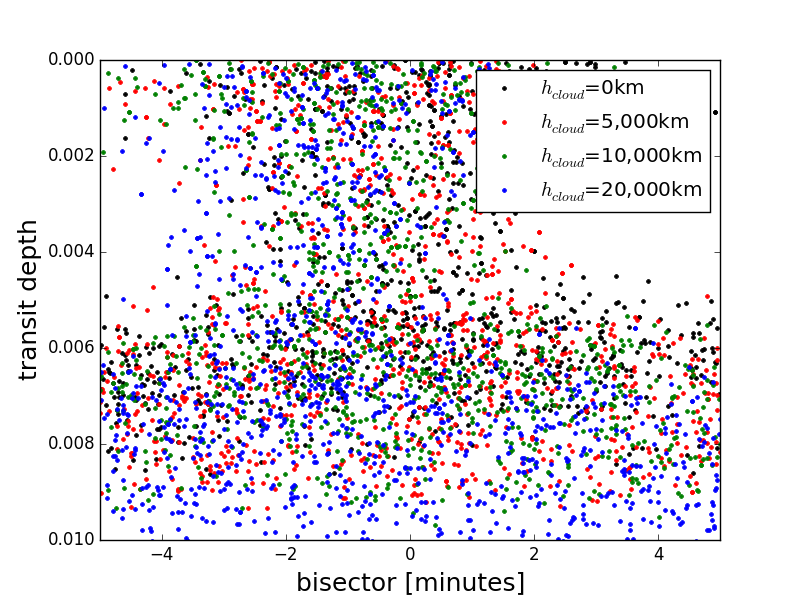}\\
\caption{Same as Fig. \ref{bisect}, but with synthetic Gaussian noise on the original light curve. Upper panel: 34\,ppm. Lower panel: 500\,ppm. Note the differences in horizontal scale.}
\label{bisect_noise}
\end{center}
\end{figure}

The bisector is strongly affected by the observational noise at high transit depths, where the light curve is flat. Hence small variations due to noise will drastically change the bisector span.  Figure \ref{bisect_noise} (upper panel) shows the same bisector spans, but with a synthetic noise of 34\,ppm added to the light curve. This is the expected noise for stars of magnitude 8 with PlaTO 2.0 \citep{rauer2014}. The bisector variations with transit depth are still clearly visible.

When adding a more realistic noise with an amplitude of 500\,ppm, comparable to Kepler magnitude 11-12 stars, the shape of the bisector is hard to analyze. It may be possible to exclude high cloud altitudes of the order of 20,000\,km, but note, however, that such an altitude would correspond to about 13 scale heights in the case of Kepler-7b. 

\begin{figure}[h]
\begin{center}
\includegraphics[width=250pt]{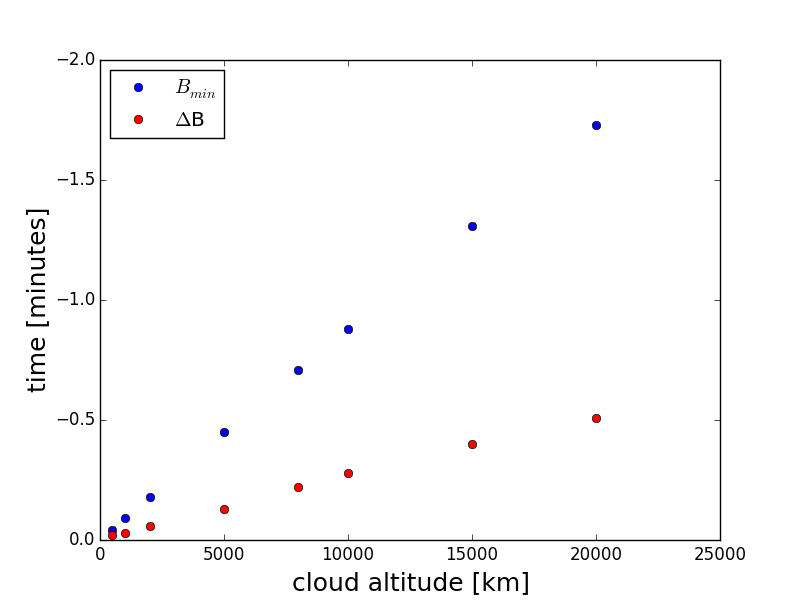}\\
\caption{Bisector diagnostics as a function of cloud altitude for the same setup as in Fig. \ref{bisect} (see text for discussion).}
\label{bisect_diag}
\end{center}
\end{figure}

Two main characteristics are proposed to analyze the bisector, as a function of cloud altitude. These are, firstly, the minimum bisector at half ingress (when the entire leading hemisphere already covers the star and the trailing hemisphere begins covering the star), $B_{\rm{min}}$. Since an offset in $\Phi_0$ of a symmetric transit also leads to a non-zero $B_{\rm{min}}$, the second criterion evaluates the shape of the bisector. It uses the difference $\Delta B$ between the bisector at the beginning of the ingress and at half ingress. Figure \ref{bisect_diag} shows these two diagnostics as a function of cloud altitude for a Kepler-7 setup. It is clearly seen that there is an almost linear relationship between the diagnostics and cloud altitude. The slope of the relationship depends on stellar characteristics (limb darkening) and the planetary orbit (especially $a_S$).

To develop a detection criterion, the rms of the bisector $\sigma_B$ is calculated as the standard deviation over ingress and egress. This excludes the full-transit phase where the planet covers the star completely and the bisector is most sensitive to observational noise. Then, the median $B_{\rm{med}}$ over the same range is calculated to establish a non-zero bisector:

\begin{equation}
\label{nonzerodetect}
\sigma_0=\frac{B_{\rm{med}}}{\sigma_B}.
\end{equation}

Once a non-zero bisector has been established, the shape is analyzed with the following criterion on a smoothed bisector:

\begin{equation}
\label{shapedetect}
\sigma_{\rm{shape}}=\frac{\Delta B}{\sigma_B}.
\end{equation}

For the 34\,ppm cases in Fig. \ref{bisect_noise}, we thus find significant detections ($\sigma_0$ between 2 and 6, and subsequently, $\sigma_{\rm{shape}}$ between 3 and 5). However, in the 500\,ppm cases, all but the 20,000\,km bisector span are compatible with zero. Even the most favorable case only shows a $\sigma_0$ of the order of 1.5, and when smoothing the bisector, one finds a $\sigma_{\rm{shape}}$ around 4. This would imply a somewhat significant detection, but as mentioned above, corresponding to much more than 10 scale heights.

\subsection{Residuals}

\begin{figure}[h]
\begin{center}
\includegraphics[width=250pt]{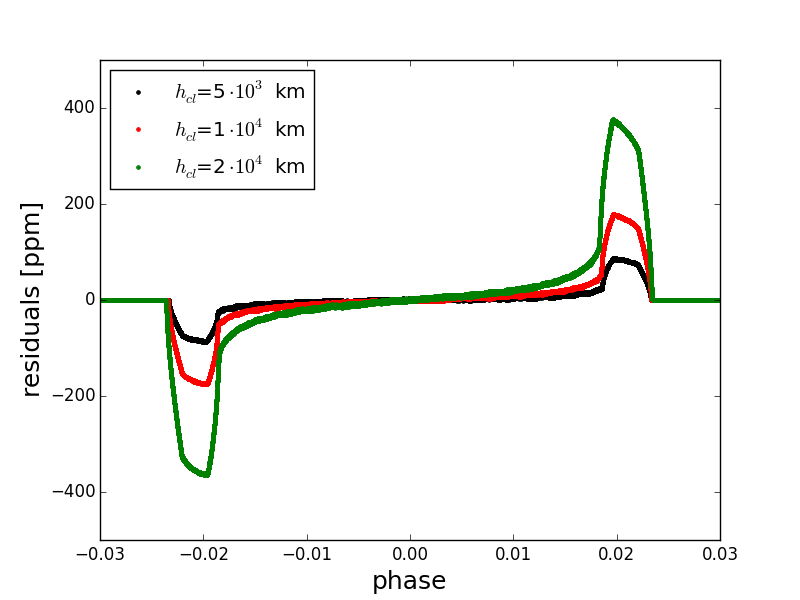}\\
\caption{Illustration of residuals between symmetric and asymmetric models.}
\label{residu_illu}
\end{center}
\end{figure}

The second cloud diagnostics are the residuals of the data with respect to a symmetric transit model, such as the model of \citet{mandel2002}. An asymmetric transit will imprint a clear structure on the residuals during ingress and egress when compared to a symmetric model. Figure \ref{residu_illu} shows the residuals between symmetric and asymmetric models, again, as above, for Kepler-7 system parameters. We compare asymmetric models with $R_P$=18\,$R_{\oplus}$ and varying cloud altitudes to symmetric models with a planetary radius $R_{\rm{eq}}$ such that the asymmetric planet and the symmetric planet cover the same surface area:

\begin{equation}
\label{req_def}
R_{\rm{eq}}^2=\frac{1}{2}\left(R_P^2+(R_P+h_{\rm{cloud}})^2\right).
\end{equation}

\begin{figure}[h]
\begin{center}
\includegraphics[width=250pt]{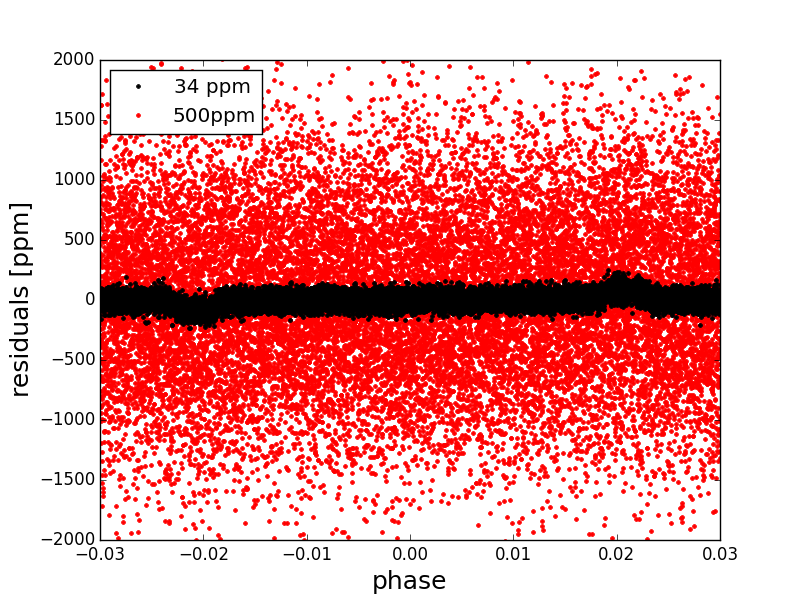}\\
\caption{Illustration of residuals between symmetric and asymmetric models with a cloud altitude of 10,000\,km: noise levels of 34 and 500\,ppm, respectively.}
\label{residu_illu_noise}
\end{center}
\end{figure}

It is clearly seen that in the short ingress and egress phases of the light curve, the residuals show distinct structures. As above for the bisector span, these residuals are somewhat sensitive to the observational noise. As illustrated in Fig. \ref{residu_illu_noise}, for a synthetic Gaussian noise of 34\,ppm, the residuals still clearly show a structure. Upon increasing the noise, however, even at the adopted cloud altitude of 10,000\,km, the residuals do not show a sign of clouds anymore.

\subsection{Degeneracies}

Equation \ref{req_def} shows the principal correlation between the parameters of the asymmetric model: in the absence of limb darkening, asymmetric and symmetric models will be identical during transit, if eq. \ref{req_def} is satisfied. If limb darkening is present, small differences between various parameter combinations allows for resolving these degeneracies. However, observational noise will tend to counteract this effect. Therefore, in the presence of significant noise, strong correlation between parameters is expected.

Combinations of planetary radius and cloud altitude that satisfy eq. \ref{req_def}, however, will show slight timing variations as to the start and end of the transit ingress and egress phases. These different starting/ending times of the planetary transit can be compensated for by choice of $\Phi_0$, thus leading to the second possible correlation between model parameters if the light curve is significantly affected by noise.

\begin{figure}[h]
\begin{center}
\includegraphics[width=250pt]{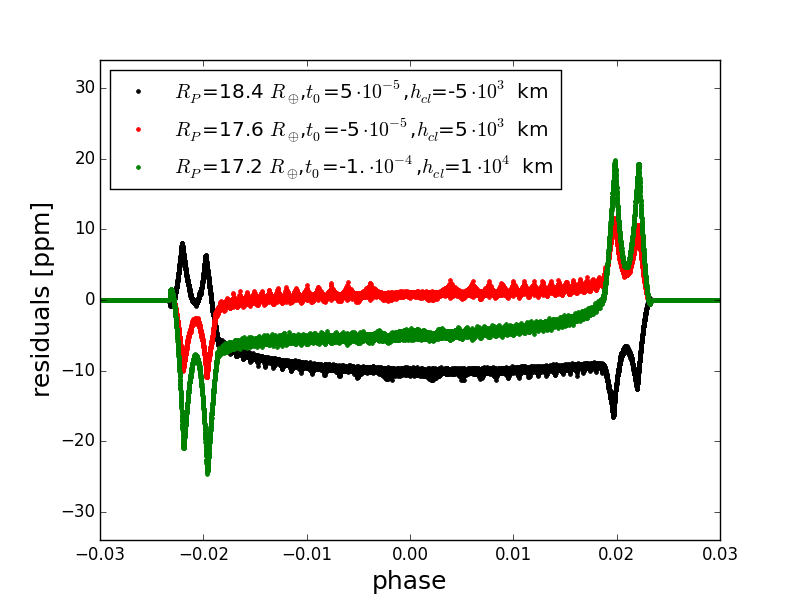}\\
\caption{Illustration of residuals between various asymmetric models and a symmetric model ($\Phi_0$=0, $h_{\rm{cloud}}$=0).}
\label{degen_illu}
\end{center}
\end{figure}

In Fig. \ref{degen_illu}, several residuals are shown with respect to a symmetric model. Note in particular the smaller vertical scale compared to Figs. \ref{residu_illu} and \ref{residu_illu_noise}.

Adding synthetic white noise to the model light curves used in Fig. \ref{degen_illu}, we were able to determine the maximum noise level, up to which a clear distinction between models, hence decorrelating the parameters, would still be possible. In general, a noise level of lower than about 50-100\,ppm allows to clearly retrieve the original model.

\section{Inverse model}

\label{invmodel}

We use the Bayesian formalism to calculate posterior probability values $p(V_P | D)$ for the parameter vector $V_P$ in the model, given a set $D$ of observations. 

\begin{equation}
\label{bayes}
p(V_P | D)\propto p(D | V_P) \cdot p(V_P).
\end{equation}

The likelihood $p(D | V_P)$ is calculated assuming independent measurements and identically-distributed Gaussian errors for the individual data points. The priors $p(V_P) $ are taken to be uninformative over the entire parameter range allowed.

\subsection{MCMC algorithm}

To sample the full parameter space, we adopt a Markov Chain Monte Carlo (MCMC) approach. We use the emcee python package developed by \citet{foreman2013}, which implements an algorithm described in \citet{goodman2010}. emcee uses multiple chains (in this work, typically around 500) to sample the parameter space. In each step, for each chain, the algorithm proposes a new positions based on the position of the entire ensemble of chains. Compared to more traditional MCMC approaches such as Metropolis-Hastings, emcee converges much quicker and is less dependent on initial conditions. Furthermore, emcee does not need a tuning of the parameter of the proposal function.

\subsection{Fitting procedure}

We proceeded in two different steps for the MCMC simulations.

In a first step, a low numerical resolution in the forward model was used, that is however sufficient to constrain stellar and orbital characteristics (i.e., $a_S$, $b$, $u_1$, $u_2$) to a high degree of confidence. To ensure good convergence and avoid any contamination by initial conditions, the chains were run for 500 steps ($>$5 auto-correlation lengths for each parameter). The first 1-2 auto-correlation lengths were considered as burn-in and discarded for the calculation of parameter uncertainties. Convergence was checked by inspecting visually the evolution of the mean of the entire ensemble. Initial positions were obtained with a random sample within the assumed prior to allow the sampler to start by exploring the entire parameter space. Uncertainty ranges are calculated by marginalizing over the posterior distribution, thinned by the auto-correlation length of the particular parameter in question. We then determine 68\,\% and 95\,\% credibility regions as the [0.16,0.84] and [0.03,0.98] median-centered percentiles, respectively, of the cumulative probability distributions (CDF). If the parameter distribution were to be Gaussian, these credibility regions would correspond to the 1\, and 2\,$\sigma$ uncertainties, respectively. 

Then, in a second step, the 95\,\% credibility regions obtained in the first step were used to initialize a high-resolution forward model. The resolution is increased such that the stellar disk is resolved better than a planetary scale height, and thus, the model becomes sensitive to the effect of clouds on the transit shape. However, this increases the computational cost by about a factor of 100.

\subsection{Scenarios}

For each planet, we considered three scenarios. The first scenario (five parameters) is a symmetric scenario (no clouds) with $\Phi_0$ fixed at 0. The second scenario relaxes this constraint and fits additionally for $\Phi_0$, but retains the assumption of no clouds, hence symmetric transits. The third scenario is fully asymmetric, fitting for all seven parameters in the model.

\section{Results}

\label{results}

\subsection{Bisector spans}

\begin{figure}[h]
\begin{center}
\includegraphics[width=250pt]{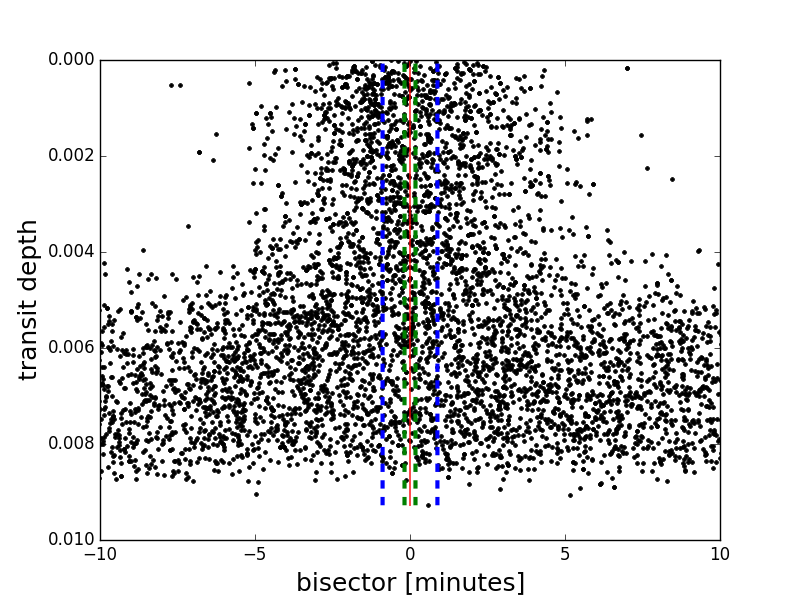}\\
\caption{Kepler-7b bisector (Q3-Q8). Red horizontal line indicates zero. Green and blue dashed lines represent $\pm$1 and $\pm$5$T_H$, respectively (see text).}
\label{bisect_k7}
\end{center}
\end{figure}

\begin{figure}[h]
\begin{center}
\includegraphics[width=250pt]{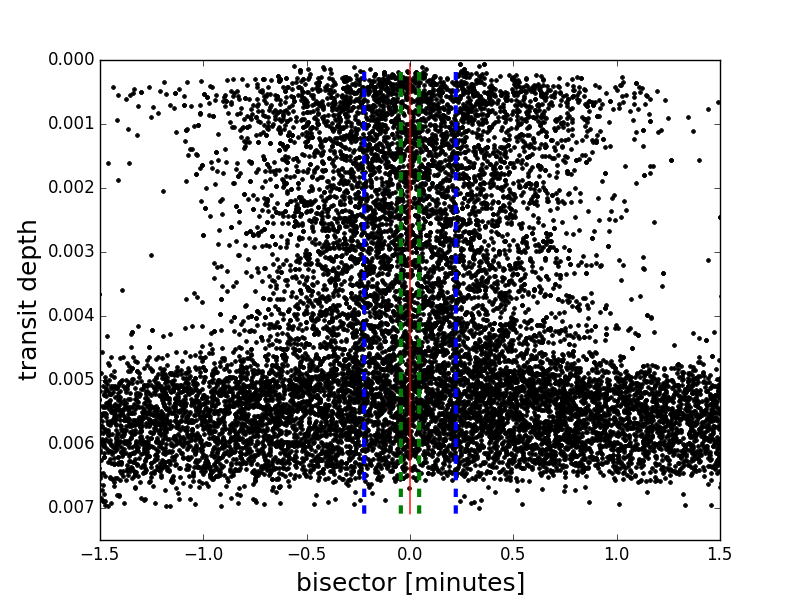}\\
\caption{HAT-P-7b bisector (Q0-Q17). Red horizontal line indicates zero. Green and blue dashed lines represent $\pm$1 and $\pm$5$T_H$, respectively. Note the smaller horizontal scale compared to Fig. \ref{bisect_k7}.}
\label{bisect_h7}
\end{center}
\end{figure}

\begin{figure}[h]
\begin{center}
\includegraphics[width=250pt]{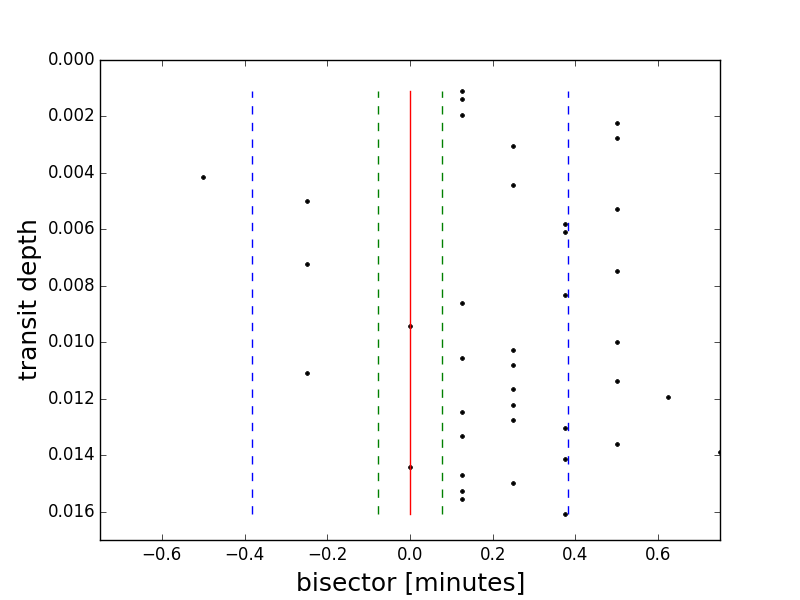}\\
\caption{HD209458b bisector. Red horizontal line indicates zero. Green and blue dashed lines represent $\pm$1 and $\pm$5$T_H$, respectively. Note the smaller horizontal scale compared to Fig. \ref{bisect_h7}.}
\label{bisect_hd}
\end{center}
\end{figure}

In Figs. \ref{bisect_k7}-\ref{bisect_hd}, the bisector spans of the three target planets are shown. As expected from the data quality in Figs. \ref{phasefoldedq3}, \ref{phasefoldedq3_hat} and \ref{trans_hd}, the bisector spans are much better constrained for HAT-P-7b and HD209458b. 

We define the scale height time, $T_H$, as the characteristic time scale on which cloud phenomena will impact the transit light curve. Essentially, it describes the time needed for the planet to advance by one scale height while transiting:
 
\begin{equation}
\label{scaletime}
T_H=\frac{H}{v_{\rm{orb}}},
\end{equation}

with $H$ the scale height and $v_{\rm{orb}}$ the orbital velocity. For Kepler-7b, we find $T_H$=10.7\,s, whereas for the relatively high-gravity planets HAT-P-7b and HD209458b, $T_H$=2.7\,s and $T_H$=4.5\,s, respectively. In Figs. \ref{bisect_k7}-\ref{bisect_hd}, we show the scale height times $T_H$ for the different planets (green lines: $\pm$$T_H$, blue lines $\pm$5$T_H$) to illustrate the potential constraints that could be inferred from the bisector spans. 

It is clear that neither Kepler-7b nor HAT-P-7b show a definitive sign of asymmetry in the bisector span. Following the detection criteria developed in eqs. \ref{nonzerodetect} and \ref{shapedetect}, we find $\sigma_{0}$ of far less than unity for both Kepler-7b and HAT-P-7b. This is because for Kepler-7b the photometric noise is too high and for HAT-P-7b, $T_H$ simply is too small. It should be noted, though, that for a Kepler-7b-like planet orbiting around a star similar to HAT-P-7 or HD209458, the bisector span could at least have been used to constrain clouds to within a few scale heights, since Kepler-7b has a very favorable scale height.

In the case of HD209458b, the bisector span is very well constrained to within a few scale heights. However, due to the low time sampling of the transit light curve of HD209458b (only four transits, covered by a few hundred points, \citealp{brown2001}), the bisector span of HD209458b is not conclusive to infer reliably the presence of any asymmetry. 

\subsection{Residuals}


In Appendix \ref{mcmcresults}, the parameter space projections of the symmetric MCMC simulations are shown (Figs. \ref{K7_triangle_no}, \ref{K7_triangle_t},  \ref{H7_triangle_no} and \ref{H7_triangle_t}). All stellar, orbital and planetary parameters are well constrained for both planets. In the case of Kepler-7b, the symmetric models show that the observations are consistent with $\Phi_0$=0, whereas for HAT-P-7b, a statistically significant deviation of the order of 2\,sec is found by the MCMC model. This is slightly larger than the 1\,$\sigma$ uncertainty associated with $T_0$ in Table \ref{dr24}. The best-fitting models have a $\chi^2_{\rm{red}}$ value of 1 and 0.99 for Kepler-7b and HAT-P-7b, respectively. Hence the quality of the fit is satisfactory.





\begin{figure}[h]
\begin{center}
\includegraphics[width=250pt]{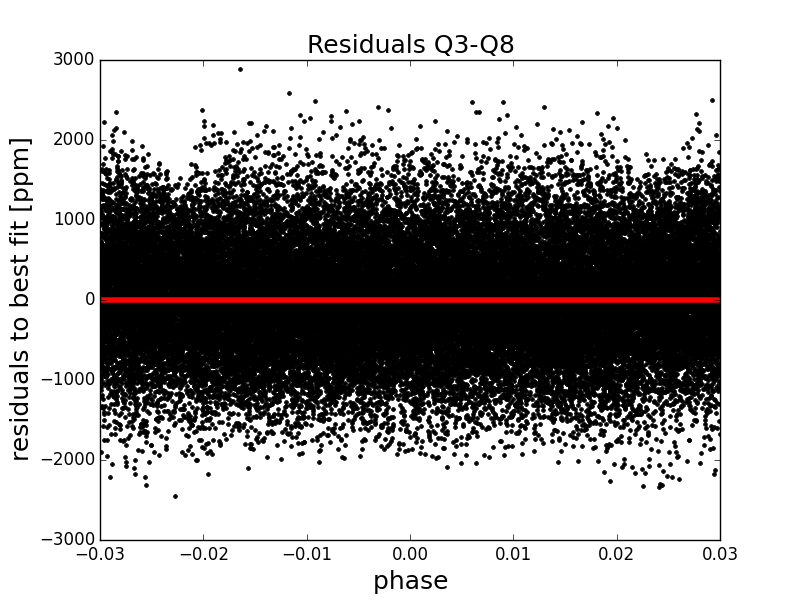}\\
\caption{Best-fit residuals for Kepler-7b.}
\label{showres}
\end{center}
\end{figure}
\begin{figure}[h]
\begin{center}
  \includegraphics[width=250pt]{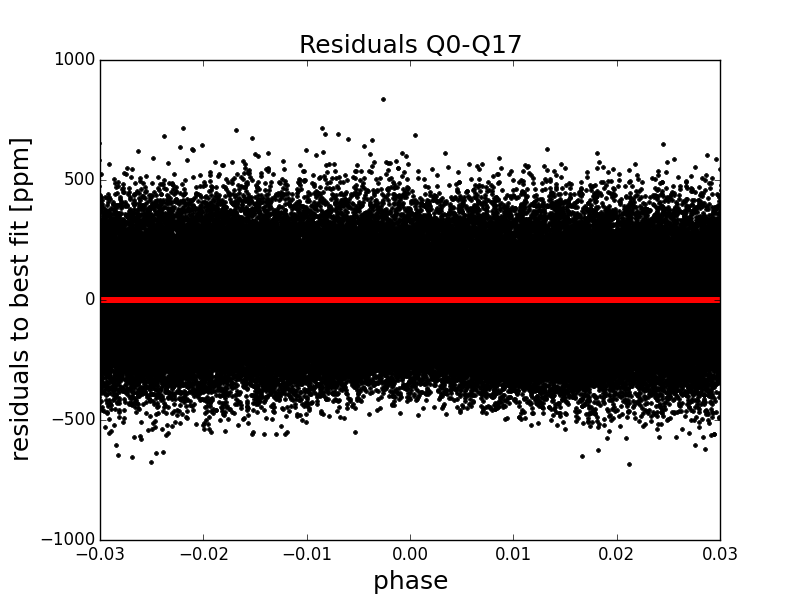}\\
\caption{Best-fit residuals for HAT-P-7b. Note the difference in vertical scale compared to Fig. \ref{showres}.}
\label{showres2}
\end{center}
\end{figure}

In Figs. \ref{showres} and \ref{showres2}, the residuals of the symmetric fits to the data are shown. As before with the bisector spans, the residuals do not show any hint of asymmetry or clouds comparable to the ones discussed above in Sect. \ref{clouddiag}.

Again, when comparing the results for both planets, it is apparent that a Kepler-7b-like planet around a relatively bright star such as HAT-P-7 would be much more favorable for such an analysis and could be amenable to cloud altitude constraints with this technique.

\subsection{Asymmetric transit models}


The parameter space projections of the asymmetric MCMC simulations are shown in Figs. \ref{K7_triangle_asym} and \ref{H7_triangle_asym}).  It is evident that in the case of Kepler-7b, no clear conclusions can be drawn from the analysis, since planetary radius $R_p$, cloud altitude $h_{\rm{cloud}}$ and phase offset $\phi_0$ are strongly correlated parameters, as expected from the high noise level (see discussion above). Hence, for Kepler-7b, the analysis is compatible with $h_{\rm{cloud}}$=0.

Note, however, that in the case of HAT-P-7b, where the standard rms of the final light curve is 150\,ppm instead of the roughly 600\,ppm for Kepler-7b, the degeneracy is somewhat weaker, and some constraints appear to be feasible. From the marginalized posterior distributions, it seems that a slight asymmetry is preferred, with $h_{\rm{cloud}}$<0 at 68\,\% confidence (although $h_{\rm{cloud}}$=0 is within the 95\,\% credibility region). This would suggest that the trailing limb appears slightly larger (of the order of a couple of scale heights). Such a result is actually consistent with the analysis of \citet{vparis2015phase} who found that the asymmetric phase curve of HAT-P-7b is more likely due to an asymmetry in scattering properties than due to a thermal offset, contrary to previous studies (e.g., \citealp{esteves2015}). However, given that the constraints are very weak, such a conclusion appears to be only tentative.

Note, in addition, that the asymmetric model contains more parameters than the symmetric models. Therefore, we apply the Bayesian Information Criterion (BIC) to choose between models. The HAT-P-7b asymmetric model is preferred by a factor of about 400 over the purely symmetric model. However, the $\chi^2$ of both the asymmetric model and the free-$\Phi_0$ model are virtually identical, so this model in turn is preferred by about a factor of 400 over the asymmetric model ($\Delta$BIC of about 12, respectively). Table \ref{mcmcsum} summarizes the fit results for the two planets and shows the planetary parameters (orbital and stellar parameters are virtually identical for all fit scenarios).

\begin{table*}
  \centering 
  \caption{95\,\% credibility regions of the planetary parameters ($R_P$, $h_{\rm{cloud}}$, $\Phi_0$) of our model fits. $\Delta$BIC and model probability ratios $p_M$ with respect to best model are stated (best model in bold).}\label{mcmcsum}
  \begin{tabular}{l|llcc|lll}
\hline
\hline
   planet& scenario & $R_P$ [$R_J$]& $\Phi_0$ [min]& $h_{\rm{cloud}}$ [km] &  $\Delta$ BIC & $p_M$\\
\hline
\hline
Kepler-7b  &   &&&\\
  & \textbf{symmetric}  & 17.92<$R_P$<18.04&-&-&0&1\\
  & symmetric + $\Phi_0$  &17.92<$R_P$<18.04&-0.12<$\Phi_0$<0.05&-&-10.2&$6.0 \cdot 10^{-3}$\\
  & asymmetric  &16.74<$R_P$<18.91&-1.17<$\Phi_0$<0.90& -12,450<$h_{\rm{cloud}}$<15,370&-20.9&$2.8 \cdot 10^{-5}$\\
\hline
\hline
HAT-P-7b  &  && &&&\\
  & symmetric  &16.09<$R_P$<16.10&-&-&-23.9&$6.3 \cdot 10^{-6}$\\
 & \textbf{symmetric + $\Phi_0$}  &16.09<$R_P$<16.10&-0.03<$\Phi_0$<-0.01&- & 0&1\\
  & asymmetric  &15.98<$R_P$<16.39&-0.10<$\Phi_0$<0.19&-3,860<$h_{\rm{cloud}}$<1,370&-12.0& $2.4 \cdot 10^{-3}$\\
\end{tabular}
\end{table*}

\section{Discussion}

\label{discuss}

\subsection{Suitable targets}

Although it affects the entire transit light curve via limb darkening, most of the effect of the phenomenon that is looked for occurs on extremely short timescales, of the order of a few to a few tens of seconds. Therefore, the light curve must be obtained at a very high cadence. This can be done with short exposure times or with a long time series covering dozens, or even hundreds, of orbits. Furthermore, bright targets are needed to obtain the required photometric precision of the order of a few tens of ppm.

In addition to the requirements for the star, the planets must also be suitable for this technique to be effective. Orbital speeds for close-in planets are of the order of 100-200\,kms$^{-1}$, hence the scale height of the considered planet should be of the order of 1,000-2,000\,km. Low-gravity planets in short orbits have very large scale heights ($H\sim \frac{T_{\rm{eq}}}{g}$, $g$ gravity, $T_{\rm{eq}}$ equilibrium temperature). For scale heights of the order of 10$^3$-10$^4$\,km, the signature of clouds during primary transit will be visible for bright targets.

\begin{figure}[h]
\begin{center}
\includegraphics[width=250pt]{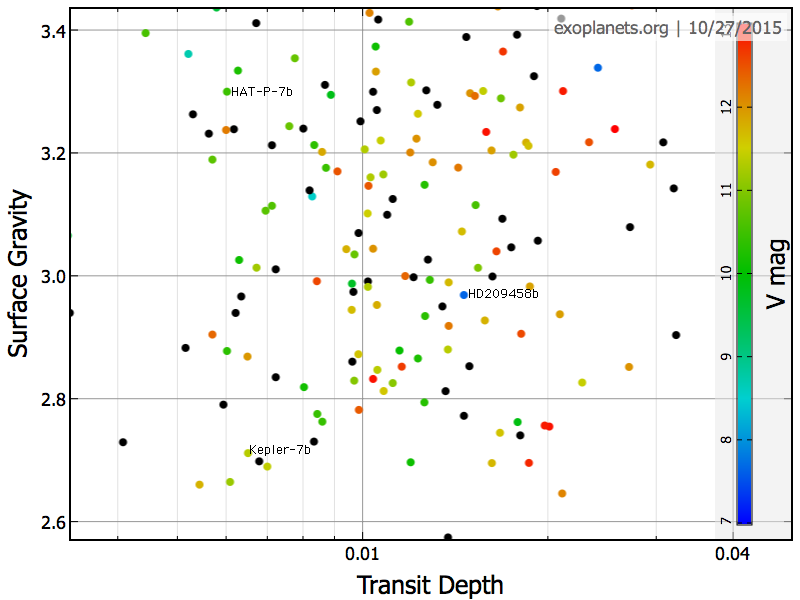}\\
\caption{Exoplanet statistics: Transit depth (x-axis) and surface gravity ($\log g$, in cm s$^{-2}$, y-axis). Color coding for the host star V magnitude. Planets used in this work are indicated. Data and plot taken from exoplanets.org (see \citealp{wright2011} and \citealp{han2014} for detailed description of the database).}
\label{targets_choice}
\end{center}
\end{figure}

Figure \ref{targets_choice} shows a compilation of transiting planets discovered so far. The host star V magnitude is used as a color coding, and we also show transit depth (x-axis) and surface gravity (y-axis). As is clear, suitable targets (in terms of the diagram, blue dots in the lower quarter of the diagram) are missing. For instance, as already noted above, HAT-P-7 is a reasonably bright target, but the surface gravity is much too large. Inversely, Kepler-7b has a very suitable surface gravity (only slightly larger than Mars), but orbits a too faint host star for this technique to work.

Many transiting planets have been discovered to date with very low surface gravities, down to $\sim$1\,ms$^{-2}$ (e.g., \citealp{hartman2011}, \citealp{lissauer2011}, \citealp{bonfils2012}, \citealp{jontof2014}). However, as for Kepler-7b, their host stars are too faint for cloud detection with primary transits.

Upcoming space missions such as ChEOPS \citep{broeg2013}, TESS \citep{ricker2014} and PlaTO 2.0 \citep{rauer2014} are expected to deliver many more transiting planets and candidates as well as high-precision photometry of radial-velocity planets, but, in contrast to CoRoT and Kepler, around much brighter host stars (down to magnitude 8). Therefore, the detection of cloud signatures via primary transit asymmetries should be feasible for at least a few candidates from these surveys.

\subsection{Forward scattering clouds}

Optically thick clouds are responsible for the reflected light contribution to the observed phase curves and are the aim of this work. However, optically thin clouds can also play a large role in shaping the transmission across the planetary limbs. Depending on cloud particle characteristics, a strong forward scattering peak can appear (e.g., \citealp{wakeford2015}), which then scatters stellar light into the line-of-sight.   Therefore, the presence of high-altitude, optically thin clouds can lead to a decreased contrast between cloud-free and cloud-covered limbs. Detailed modeling of such situations is however beyond the scope of this paper.

\subsection{Other sources of asymmetric transits}

Uniquely attributing asymmetric transit light curves to inhomogeneous cloud cover is not possible. Over the last decade, several other astrophysical sources of asymmetric transit light curves have been presented. 

For instance, 3D atmospheric modeling of hot Jupiters generally predicts strong eastwards circulation (e.g., \citealp{showman2002}), and thus, a hotter trailing limb than the morning limb (e.g., \citealp{agundez2012,agundez2014}, \citealp{parmentier2013}). It follows that the atmosphere extends further, hence the transit depth is larger.  \citet{agundez2014} note that this thermal effect dominates over the effect of chemistry in transmission spectra. In the science case considered in this work, the hotter evening side somewhat diminishes the effect of the inhomogeneous cloud cover, since the "missing" planetary radius is compensated for by the extended atmosphere on the trailing limb. On the other hand, for otherwise homogenous planets (for instance, with global cloud cover), the purely thermal effect could produce slightly asymmetric transits. 

In an order-of-magnitude estimate, when calculating the extent of the atmosphere from the homogenized high-pressure layer at $p_H$ to the cloud layer at $p_C$, one finds that the evening atmosphere is extended (compared to the morning atmosphere) by an amount $h_E$:

\begin{equation}
\label{evening}
h_E=\ln \left(\frac{p_H}{p_C}\right) \frac{\Delta T}{T_M}\cdot H,
\end{equation}

where $H$ is the scale height, $\Delta T$ is the temperature difference between morning temperature $T_M$ and evening temperature $T_E$. For many planets, 3D circulation models find relatively moderate values of about $\frac{\Delta T}{T_M} \approx$10-20\,\%. Using $p_H$=1\,bar and $p_C$=\,50mbar as representative values, one then finds $h_E\approx$0.3-0.6\,$H$. This is not negligible, however not expected to be the dominant photometric effect. 

Other possible physical causes of asymmetric transits are, e.g., the presence of exomoons around the planet in question (e.g., \citealp{szabo2006}, \citealp{kipping2009ttv}, \citealp{heller2014}) or gas and dust tails of strongly irradiated or disrupting planets (e.g., \citealp{brogi2012}, \citealp{rappaport2014}, \citealp{sanchis2015}). 

A possible way to discriminate between such scenarios would be, for instance, that in the case of tailed exoplanets, the transit depth would vary quite remarkably. Previous studies of HAT-P-7b have revealed a slight variation of the transit depth with the observation quarter \citep{eylen2013}, which however is not attributed to physical reasons, but rather to instrumental effects. Detailed model comparisons between competing scenarios are however not the subject of this work, but rather for future studies.

\section{Conclusions}

\label{summary}

We have presented a new technique to identify cloud signatures in the transit light curves of exoplanets. Inhomogeneous cloud cover produces asymmetries in the transit light curve. Such asymmetries reveal themselves via the bisector span and characteristic residuals with respect to standard, symmetric models.

We have applied this reasoning to search for hints for clouds in the transits of Kepler-7b, HAT-P-7b and HD209458b.  No clear sign of clouds have been found, although for HAT-P-7b, results seem to indicate a small, yet not statistically very significant, asymmetry consistent with previous phase curve analyses by \citet{vparis2015phase}. This is due to several reasons. In the case of Kepler-7b, the photometric noise is too high. HAT-P-7b, despite its better photometric quality, has a scale height which is too small to allow constraints on the presence of clouds or the cloud altitude. In the case of HD209458b, the time sampling of the used transit light curve is too small, which again prevents a clear detection of clouds.

However, we have shown that with future photometric surveys and suitable targets, asymmetric transits will be detectable. This will help to better constrain cloud properties on exoplanets.

\begin{acknowledgements}

This study has received financial support from the French State in the frame of the "Investments for the future" Programme IdEx Bordeaux, reference ANR-10-IDEX-03-02. Computer time for this study was provided in parts by the computing facilities MCIA (M\'esocentre de Calcul Intensif Aquitain) of the Universit\'e de Bordeaux and of the Universit\'e de Pau et des Pays de l'Adour. P. G. acknowledges support from the ERC Starting Grant (3DICE, grant agreement 336474). P. G.s postdoctoral position is funded by INSU/CNRS. This paper includes data collected by the Kepler mission. Funding for the Kepler mission is provided by the NASA Science Mission directorate. Some of the data presented in this paper were obtained from the Mikulski Archive for Space Telescopes (MAST). STScI is operated by the Association of Universities for Research in Astronomy, Inc., under NASA contract NAS5-26555. Support for MAST for non-HST data is provided by the NASA Office of Space Science via grant NNX09AF08G and by other grants and contracts. This research has made use of the Exoplanet Orbit Database
and the Exoplanet Data Explorer at exoplanets.org.

\end{acknowledgements}

\bibliographystyle{aa}
\bibliography{literatur_idex}

\begin{thebibliography}{53}
\expandafter\ifx\csname natexlab\endcsname\relax\def\natexlab#1{#1}\fi

\bibitem[{{Ag{\'u}ndez} {et~al.}(2014){Ag{\'u}ndez}, {Parmentier}, {Venot},
  {Hersant}, \& {Selsis}}]{agundez2014}
{Ag{\'u}ndez}, M., {Parmentier}, V., {Venot}, O., {Hersant}, F., \& {Selsis},
  F. 2014, \aap, 564, A73

\bibitem[{{Ag{\'u}ndez} {et~al.}(2012){Ag{\'u}ndez}, {Venot}, {Iro}, {Selsis},
  {Hersant}, {H{\'e}brard}, \& {Dobrijevic}}]{agundez2012}
{Ag{\'u}ndez}, M., {Venot}, O., {Iro}, N., {et~al.} 2012, \aap, 548, A73

\bibitem[{{Barclay} {et~al.}(2012){Barclay}, {Huber}, {Rowe}, {Fortney},
  {Morley}, {Quintana}, {Fabrycky}, {Barentsen}, {Bloemen}, {Christiansen},
  {Demory}, {Fulton}, {Jenkins}, {Mullally}, {Ragozzine}, {Seader}, {Shporer},
  {Tenenbaum}, \& {Thompson}}]{barclay2012}
{Barclay}, T., {Huber}, D., {Rowe}, J.~F., {et~al.} 2012, \apj, 761, 53

\bibitem[{{Bean} {et~al.}(2010){Bean}, {Kempton}, \& {Homeier}}]{bean2010}
{Bean}, J.~L., {Kempton}, E., \& {Homeier}, D. 2010, \nat, 468, 669

\bibitem[{{Bonfils} {et~al.}(2012){Bonfils}, {Gillon}, {Udry}, {Armstrong},
  {Bouchy}, {Delfosse}, {Forveille}, {Fumel}, {Jehin}, {Lendl}, {Lovis},
  {Mayor}, {McCormac}, {Neves}, {Pepe}, {Perrier}, {Pollaco}, {Queloz}, \&
  {Santos}}]{bonfils2012}
{Bonfils}, X., {Gillon}, M., {Udry}, S., {et~al.} 2012, \aap, 546, A27

\bibitem[{{Borucki} {et~al.}(2009){Borucki}, {Koch}, {Jenkins}, {Sasselov},
  {Gilliland}, {Batalha}, {Latham}, {Caldwell}, {Basri}, {Brown},
  {Christensen-Dalsgaard}, {Cochran}, {DeVore}, {Dunham}, {Dupree}, {Gautier},
  {Geary}, {Gould}, {Howell}, {Kjeldsen}, {Lissauer}, {Marcy}, {Meibom},
  {Morrison}, \& {Tarter}}]{borucki2009}
{Borucki}, W.~J., {Koch}, D., {Jenkins}, J., {et~al.} 2009, Science, 325, 709

\bibitem[{{Broeg} {et~al.}(2013){Broeg}, {Fortier}, {Ehrenreich}, {Alibert},
  {Baumjohann}, {Benz}, {Deleuil}, {Gillon}, {Ivanov}, {Liseau}, {Meyer},
  {Oloffson}, {Pagano}, {Piotto}, {Pollacco}, {Queloz}, {Ragazzoni}, {Renotte},
  {Steller}, \& {Thomas}}]{broeg2013}
{Broeg}, C., {Fortier}, A., {Ehrenreich}, D., {et~al.} 2013, in European
  Physical Journal Web of Conferences, Vol.~47, European Physical Journal Web
  of Conferences, 3005

\bibitem[{{Brogi} {et~al.}(2012){Brogi}, {Keller}, {de Juan Ovelar},
  {Kenworthy}, {de Kok}, {Min}, \& {Snellen}}]{brogi2012}
{Brogi}, M., {Keller}, C.~U., {de Juan Ovelar}, M., {et~al.} 2012, \aap, 545,
  L5

\bibitem[{{Brown} {et~al.}(2001){Brown}, {Charbonneau}, {Gilliland}, {Noyes},
  \& {Burrows}}]{brown2001}
{Brown}, T.~M., {Charbonneau}, D., {Gilliland}, R.~L., {Noyes}, R.~W., \&
  {Burrows}, A. 2001, \apj, 552, 699

\bibitem[{{Deming} {et~al.}(2013){Deming}, {Wilkins}, {McCullough}, {Burrows},
  {Fortney}, {Agol}, {Dobbs-Dixon}, {Madhusudhan}, {Crouzet}, {Desert},
  {Gilliland}, {Haynes}, {Knutson}, {Line}, {Magic}, {Mandell}, {Ranjan},
  {Charbonneau}, {Clampin}, {Seager}, \& {Showman}}]{deming2013}
{Deming}, D., {Wilkins}, A., {McCullough}, P., {et~al.} 2013, \apj, 774, 95

\bibitem[{{Demory} {et~al.}(2013){Demory}, {de Wit}, {Lewis}, {Fortney},
  {Zsom}, {Seager}, {Knutson}, {Heng}, {Madhusudhan}, {Gillon}, {Barclay},
  {Desert}, {Parmentier}, \& {Cowan}}]{demory2013inhomogen}
{Demory}, B.-O., {de Wit}, J., {Lewis}, N., {et~al.} 2013, \apjl, 776, L25

\bibitem[{{Demory} {et~al.}(2011){Demory}, {Seager}, {Madhusudhan}, {Kjeldsen},
  {Christensen-Dalsgaard}, {Gillon}, {Rowe}, {Welsh}, {Adams}, {Dupree},
  {McCarthy}, {Kulesa}, {Borucki}, \& {Koch}}]{demory2011kepler7}
{Demory}, B.-O., {Seager}, S., {Madhusudhan}, N., {et~al.} 2011, \apjl, 735,
  L12

\bibitem[{{Esteves} {et~al.}(2013){Esteves}, {De Mooij}, \&
  {Jayawardhana}}]{esteves2013}
{Esteves}, L.~J., {De Mooij}, E.~J.~W., \& {Jayawardhana}, R. 2013, \apj, 772,
  51

\bibitem[{{Esteves} {et~al.}(2015){Esteves}, {De Mooij}, \&
  {Jayawardhana}}]{esteves2015}
{Esteves}, L.~J., {De Mooij}, E.~J.~W., \& {Jayawardhana}, R. 2015, \apj, 804,
  150

\bibitem[{{Evans} {et~al.}(2013){Evans}, {Pont}, {Sing}, {Aigrain}, {Barstow},
  {D{\'e}sert}, {Gibson}, {Heng}, {Knutson}, \& {Lecavelier des
  Etangs}}]{evans2013}
{Evans}, T.~M., {Pont}, F., {Sing}, D.~K., {et~al.} 2013, \apjl, 772, L16

\bibitem[{{Foreman-Mackey} {et~al.}(2013){Foreman-Mackey}, {Hogg}, {Lang}, \&
  {Goodman}}]{foreman2013}
{Foreman-Mackey}, D., {Hogg}, D.~W., {Lang}, D., \& {Goodman}, J. 2013, \pasp,
  125, 306

\bibitem[{{Gibson} {et~al.}(2013){Gibson}, {Aigrain}, {Barstow}, {Evans},
  {Fletcher}, \& {Irwin}}]{gibson2013}
{Gibson}, N.~P., {Aigrain}, S., {Barstow}, J.~K., {et~al.} 2013, \mnras, 436,
  2974

\bibitem[{{Goodman} \& {Weare}(2010)}]{goodman2010}
{Goodman}, J. \& {Weare}, J. 2010, Commun. Appl. Math. Comput. Sci., 5

\bibitem[{{Han} {et~al.}(2014){Han}, {Wang}, {Wright}, {Feng}, {Zhao},
  {Fakhouri}, {Brown}, \& {Hancock}}]{han2014}
{Han}, E., {Wang}, S.~X., {Wright}, J.~T., {et~al.} 2014, \pasp, 126, 827

\bibitem[{{Hansen} {et~al.}(2014){Hansen}, {Schwartz}, \& {Cowan}}]{hansen2014}
{Hansen}, C.~J., {Schwartz}, J.~C., \& {Cowan}, N.~B. 2014, \mnras, 444, 3632

\bibitem[{{Hartman} {et~al.}(2011){Hartman}, {Bakos}, {Kipping}, {Torres},
  {Kov{\'a}cs}, {Noyes}, {Latham}, {Howard}, {Fischer}, {Johnson}, {Marcy},
  {Isaacson}, {Quinn}, {Buchhave}, {B{\'e}ky}, {Sasselov}, {Stefanik},
  {Esquerdo}, {Everett}, {Perumpilly}, {L{\'a}z{\'a}r}, {Papp}, \&
  {S{\'a}ri}}]{hartman2011}
{Hartman}, J.~D., {Bakos}, G.~{\'A}., {Kipping}, D.~M., {et~al.} 2011, \apj,
  728, 138

\bibitem[{{Heller}(2014)}]{heller2014}
{Heller}, R. 2014, \apj, 787, 14

\bibitem[{{Jontof-Hutter} {et~al.}(2014){Jontof-Hutter}, {Lissauer}, {Rowe}, \&
  {Fabrycky}}]{jontof2014}
{Jontof-Hutter}, D., {Lissauer}, J.~J., {Rowe}, J.~F., \& {Fabrycky}, D.~C.
  2014, \apj, 785, 15

\bibitem[{{Kipping}(2009)}]{kipping2009ttv}
{Kipping}, D.~M. 2009, \mnras, 392, 181

\bibitem[{{Knutson} {et~al.}(2014){Knutson}, {Benneke}, {Deming}, \&
  {Homeier}}]{knutson2014}
{Knutson}, H.~A., {Benneke}, B., {Deming}, D., \& {Homeier}, D. 2014, \nat,
  505, 66

\bibitem[{{Kreidberg} {et~al.}(2014){Kreidberg}, {Bean}, {D{\'e}sert},
  {Benneke}, {Deming}, {Stevenson}, {Seager}, {Berta-Thompson}, {Seifahrt}, \&
  {Homeier}}]{kreidberg2014}
{Kreidberg}, L., {Bean}, J.~L., {D{\'e}sert}, J.-M., {et~al.} 2014, \nat, 505,
  69

\bibitem[{{Latham} {et~al.}(2010){Latham}, {Borucki}, {Koch}, {Brown},
  {Buchhave}, {Basri}, {Batalha}, {Caldwell}, {Cochran}, {Dunham}, {F{\H
  u}r{\'e}sz}, {Gautier}, {Geary}, {Gilliland}, {Howell}, {Jenkins},
  {Lissauer}, {Marcy}, {Monet}, {Rowe}, \& {Sasselov}}]{latham2010}
{Latham}, D.~W., {Borucki}, W.~J., {Koch}, D.~G., {et~al.} 2010, \apjl, 713,
  L140

\bibitem[{{Lee} {et~al.}(2015){Lee}, {Helling}, {Dobbs-Dixon}, \&
  {Juncher}}]{lee2015}
{Lee}, G., {Helling}, C., {Dobbs-Dixon}, I., \& {Juncher}, D. 2015, \aap, 580,
  A12

\bibitem[{{Line} \& {Parmentier}(2015)}]{line2015}
{Line}, M.~R. \& {Parmentier}, V. 2015, submitted to \apj
  [\eprint[arXiv]{1511.09443}]

\bibitem[{{Lissauer} {et~al.}(2011){Lissauer}, {Fabrycky}, {Ford}, {Borucki},
  {Fressin}, {Marcy}, {Orosz}, {Rowe}, {Torres}, {Welsh}, {Batalha}, {Bryson},
  {Buchhave}, {Caldwell}, {Carter}, {Charbonneau}, {Christiansen}, {Cochran},
  {Desert}, {Dunham}, {Fanelli}, {Fortney}, {Gautier}, {Geary}, {Gilliland},
  {Haas}, {Hall}, {Holman}, {Koch}, {Latham}, {Lopez}, {McCauliff}, {Miller},
  {Morehead}, {Quintana}, {Ragozzine}, {Sasselov}, {Short}, \&
  {Steffen}}]{lissauer2011}
{Lissauer}, J.~J., {Fabrycky}, D.~C., {Ford}, E.~B., {et~al.} 2011, \nat, 470,
  53

\bibitem[{{Mandel} \& {Agol}(2002)}]{mandel2002}
{Mandel}, K. \& {Agol}, E. 2002, \apjl, 580, L171

\bibitem[{{Mazeh} \& {Faigler}(2010)}]{mazeh2010}
{Mazeh}, T. \& {Faigler}, S. 2010, \aap, 521, L59

\bibitem[{{P{\'a}l} {et~al.}(2008){P{\'a}l}, {Bakos}, {Torres}, {Noyes},
  {Latham}, {Kov{\'a}cs}, {Marcy}, {Fischer}, {Butler}, {Sasselov}, {Sip{\H
  o}cz}, {Esquerdo}, {Kov{\'a}cs}, {Stefanik}, {L{\'a}z{\'a}r}, {Papp}, \&
  {S{\'a}ri}}]{pal2008}
{P{\'a}l}, A., {Bakos}, G.~{\'A}., {Torres}, G., {et~al.} 2008, \apj, 680, 1450

\bibitem[{{Parmentier} {et~al.}(2013){Parmentier}, {Showman}, \&
  {Lian}}]{parmentier2013}
{Parmentier}, V., {Showman}, A.~P., \& {Lian}, Y. 2013, \aap, 558, A91

\bibitem[{{Quintana} {et~al.}(2013){Quintana}, {Rowe}, {Barclay}, {Howell},
  {Ciardi}, {Demory}, {Caldwell}, {Borucki}, {Christiansen}, {Jenkins},
  {Klaus}, {Fulton}, {Morris}, {Sanderfer}, {Shporer}, {Smith}, {Still}, \&
  {Thompson}}]{quintana2013}
{Quintana}, E.~V., {Rowe}, J.~F., {Barclay}, T., {et~al.} 2013, \apj, 767, 137

\bibitem[{{Rappaport} {et~al.}(2014){Rappaport}, {Barclay}, {DeVore}, {Rowe},
  {Sanchis-Ojeda}, \& {Still}}]{rappaport2014}
{Rappaport}, S., {Barclay}, T., {DeVore}, J., {et~al.} 2014, \apj, 784, 40

\bibitem[{{Rauer} {et~al.}(2014){Rauer}, {Catala}, {Aerts}, {Appourchaux},
  {Benz}, {Brandeker}, {Christensen-Dalsgaard}, {Deleuil}, {Gizon}, {Goupil},
  {G{\"u}del}, {Janot-Pacheco}, {Mas-Hesse}, {Pagano}, {Piotto}, {Pollacco},
  {Santos}, {Smith}, {Su{\'a}rez}, {Szab{\'o}}, {Udry}, {Adibekyan}, {Alibert},
  {Almenara}, {Amaro-Seoane}, {Eiff}, {Asplund}, {Antonello}, {Barnes},
  {Baudin}, {Belkacem}, {Bergemann}, {Bihain}, {Birch}, {Bonfils}, {Boisse},
  {Bonomo}, {Borsa}, {Brand{\~a}o}, {Brocato}, {Brun}, {Burleigh}, {Burston},
  {Cabrera}, {Cassisi}, {Chaplin}, {Charpinet}, {Chiappini}, {Church},
  {Csizmadia}, {Cunha}, {Damasso}, {Davies}, {Deeg}, {D{\'{\i}}az}, {Dreizler},
  {Dreyer}, {Eggenberger}, {Ehrenreich}, {Eigm{\"u}ller}, {Erikson}, {Farmer},
  {Feltzing}, {de Oliveira Fialho}, {Figueira}, {Forveille}, {Fridlund},
  {Garc{\'{\i}}a}, {Giommi}, {Giuffrida}, {Godolt}, {Gomes da Silva},
  {Granzer}, {Grenfell}, {Grotsch-Noels}, {G{\"u}nther}, {Haswell}, {Hatzes},
  {H{\'e}brard}, {Hekker}, {Helled}, {Heng}, {Jenkins}, {Johansen},
  {Khodachenko}, {Kislyakova}, {Kley}, {Kolb}, {Krivova}, {Kupka}, {Lammer},
  {Lanza}, {Lebreton}, {Magrin}, {Marcos-Arenal}, {Marrese}, {Marques},
  {Martins}, {Mathis}, {Mathur}, {Messina}, {Miglio}, {Montalban}, {Montalto},
  {Monteiro}, {Moradi}, {Moravveji}, {Mordasini}, {Morel}, {Mortier},
  {Nascimbeni}, {Nelson}, {Nielsen}, {Noack}, {Norton}, {Ofir}, {Oshagh},
  {Ouazzani}, {P{\'a}pics}, {Parro}, {Petit}, {Plez}, {Poretti}, {Quirrenbach},
  {Ragazzoni}, {Raimondo}, {Rainer}, {Reese}, {Redmer}, {Reffert},
  {Rojas-Ayala}, {Roxburgh}, {Salmon}, {Santerne}, {Schneider}, {Schou},
  {Schuh}, {Schunker}, {Silva-Valio}, {Silvotti}, {Skillen}, {Snellen}, {Sohl},
  {Sousa}, {Sozzetti}, {Stello}, {Strassmeier}, {{\v S}vanda}, {Szab{\'o}},
  {Tkachenko}, {Valencia}, {Van Grootel}, {Vauclair}, {Ventura}, {Wagner},
  {Walton}, {Weingrill}, {Werner}, {Wheatley}, \& {Zwintz}}]{rauer2014}
{Rauer}, H., {Catala}, C., {Aerts}, C., {et~al.} 2014, Experimental Astronomy,
  38, 249

\bibitem[{{Ricker} {et~al.}(2014){Ricker}, {Winn}, {Vanderspek}, {Latham},
  {Bakos}, {Bean}, {Berta-Thompson}, {Brown}, {Buchhave}, {Butler}, {Butler},
  {Chaplin}, {Charbonneau}, {Christensen-Dalsgaard}, {Clampin}, {Deming},
  {Doty}, {De Lee}, {Dressing}, {Dunham}, {Endl}, {Fressin}, {Ge}, {Henning},
  {Holman}, {Howard}, {Ida}, {Jenkins}, {Jernigan}, {Johnson}, {Kaltenegger},
  {Kawai}, {Kjeldsen}, {Laughlin}, {Levine}, {Lin}, {Lissauer}, {MacQueen},
  {Marcy}, {McCullough}, {Morton}, {Narita}, {Paegert}, {Palle}, {Pepe},
  {Pepper}, {Quirrenbach}, {Rinehart}, {Sasselov}, {Sato}, {Seager},
  {Sozzetti}, {Stassun}, {Sullivan}, {Szentgyorgyi}, {Torres}, {Udry}, \&
  {Villasenor}}]{ricker2014}
{Ricker}, G.~R., {Winn}, J.~N., {Vanderspek}, R., {et~al.} 2014, SPIE
  Conference Series, 9143, 20

\bibitem[{{Sanchis-Ojeda} {et~al.}(2015){Sanchis-Ojeda}, {Rappaport},
  {Pall{\`e}}, {Delrez}, {DeVore}, {Gandolfi}, {Fukui}, {Ribas}, {Stassun},
  {Albrecht}, {Dai}, {Gaidos}, {Gillon}, {Hirano}, {Holman}, {Howard},
  {Isaacson}, {Jehin}, {Kuzuhara}, {Mann}, {Marcy}, {Miles-P{\'a}ez},
  {Monta{\~n}{\'e}s-Rodr{\'{\i}}guez}, {Murgas}, {Narita}, {Nowak}, {Onitsuka},
  {Paegert}, {Van Eylen}, {Winn}, \& {Yu}}]{sanchis2015}
{Sanchis-Ojeda}, R., {Rappaport}, S., {Pall{\`e}}, E., {et~al.} 2015, \apj,
  812, 112

\bibitem[{{Schlawin} {et~al.}(2014){Schlawin}, {Zhao}, {Teske}, \&
  {Herter}}]{schlawin2014}
{Schlawin}, E., {Zhao}, M., {Teske}, J.~K., \& {Herter}, T. 2014, \apj, 783, 5

\bibitem[{{Showman} \& {Guillot}(2002)}]{showman2002}
{Showman}, A.~P. \& {Guillot}, T. 2002, \aap, 385, 166

\bibitem[{{Sing} {et~al.}(2015){Sing}, {Wakeford}, {Showman}, {Nikolov},
  {Fortney}, {Burrows}, {Ballester}, {Deming}, {Aigrain}, {D{\'e}sert},
  {Gibson}, {Henry}, {Knutson}, {Lecavelier des Etangs}, {Pont},
  {Vidal-Madjar}, {Williamson}, \& {Wilson}}]{sing2015}
{Sing}, D.~K., {Wakeford}, H.~R., {Showman}, A.~P., {et~al.} 2015, \mnras, 446,
  2428

\bibitem[{{Snellen} {et~al.}(2009){Snellen}, {de Mooij}, \&
  {Albrecht}}]{snellen2009}
{Snellen}, I.~A.~G., {de Mooij}, E.~J.~W., \& {Albrecht}, S. 2009, \nat, 459,
  543

\bibitem[{{Sudarsky} {et~al.}(2000){Sudarsky}, {Burrows}, \&
  {Pinto}}]{sudarsky2000}
{Sudarsky}, D., {Burrows}, A., \& {Pinto}, P. 2000, \apj, 538, 885

\bibitem[{{Szab{\'o}} {et~al.}(2006){Szab{\'o}}, {Szatm{\'a}ry}, {Div{\'e}ki},
  \& {Simon}}]{szabo2006}
{Szab{\'o}}, G.~M., {Szatm{\'a}ry}, K., {Div{\'e}ki}, Z., \& {Simon}, A. 2006,
  \aap, 450, 395

\bibitem[{{Van Eylen} {et~al.}(2012){Van Eylen}, {Kjeldsen},
  {Christensen-Dalsgaard}, \& {Aerts}}]{eylen2012}
{Van Eylen}, V., {Kjeldsen}, H., {Christensen-Dalsgaard}, J., \& {Aerts}, C.
  2012, Astron. Nachrichten, 333, 1088

\bibitem[{{Van Eylen} {et~al.}(2013){Van Eylen}, {Lindholm Nielsen}, {Hinrup},
  {Tingley}, \& {Kjeldsen}}]{eylen2013}
{Van Eylen}, V., {Lindholm Nielsen}, M., {Hinrup}, B., {Tingley}, B., \&
  {Kjeldsen}, H. 2013, \apjl, 774, L19

\bibitem[{{von Paris} {et~al.}(2016){von Paris}, {Gratier}, {Bord\'e}, \&
  {Selsis}}]{vparis2015phase}
{von Paris}, P., {Gratier}, P., {Bord\'e}, P., \& {Selsis}, F. 2016, accepted
  for publication in \aap [\eprint[arXiv]{1512.04908}]

\bibitem[{{Wakeford} \& {Sing}(2015)}]{wakeford2015}
{Wakeford}, H.~R. \& {Sing}, D.~K. 2015, \aap, 573, A122

\bibitem[{{Webber} {et~al.}(2015){Webber}, {Lewis}, {Marley}, {Morley},
  {Fortney}, \& {Cahoy}}]{webber2015}
{Webber}, M.~W., {Lewis}, N.~K., {Marley}, M., {et~al.} 2015, \apj, 804, 94

\bibitem[{{Welsh} {et~al.}(2010){Welsh}, {Orosz}, {Seager}, {Fortney},
  {Jenkins}, {Rowe}, {Koch}, \& {Borucki}}]{welsh2010}
{Welsh}, W.~F., {Orosz}, J.~A., {Seager}, S., {et~al.} 2010, \apjl, 713, L145

\bibitem[{{Winn} {et~al.}(2009){Winn}, {Johnson}, {Albrecht}, {Howard},
  {Marcy}, {Crossfield}, \& {Holman}}]{winn2009}
{Winn}, J.~N., {Johnson}, J.~A., {Albrecht}, S., {et~al.} 2009, \apjl, 703, L99

\bibitem[{{Wright} {et~al.}(2011){Wright}, {Fakhouri}, {Marcy}, {Han}, {Feng},
  {Johnson}, {Howard}, {Fischer}, {Valenti}, {Anderson}, \&
  {Piskunov}}]{wright2011}
{Wright}, J.~T., {Fakhouri}, O., {Marcy}, G.~W., {et~al.} 2011, \pasp, 123, 412

\end{thebibliography}

\appendix

\section{MCMC results}

\label{mcmcresults}


%
%

\begin{figure*}
\begin{center}
\includegraphics[width=450pt]{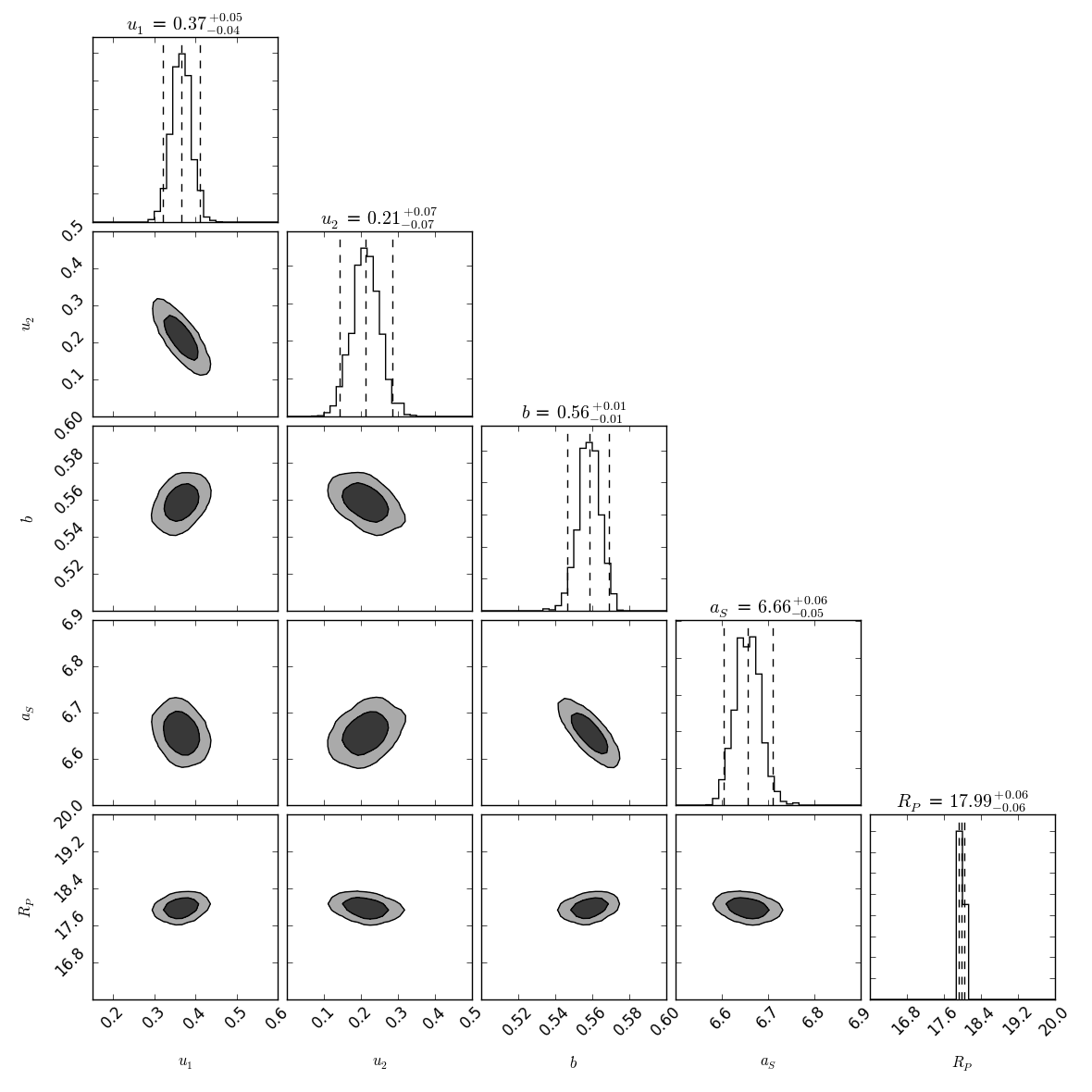}\\
\caption{Kepler-7b posterior projections for the symmetric model ($\Phi_0$=0, $h_{\rm{cloud}}$=0). Dashed vertical lines represent marginalized 95\,\% credibility regions. Smoothed 68\,\% and 95\,\% credibility regions in dark and light grey shade, respectively.}
\label{K7_triangle_no}
\end{center}
\end{figure*}

\begin{figure*}
\begin{center}
\includegraphics[width=450pt]{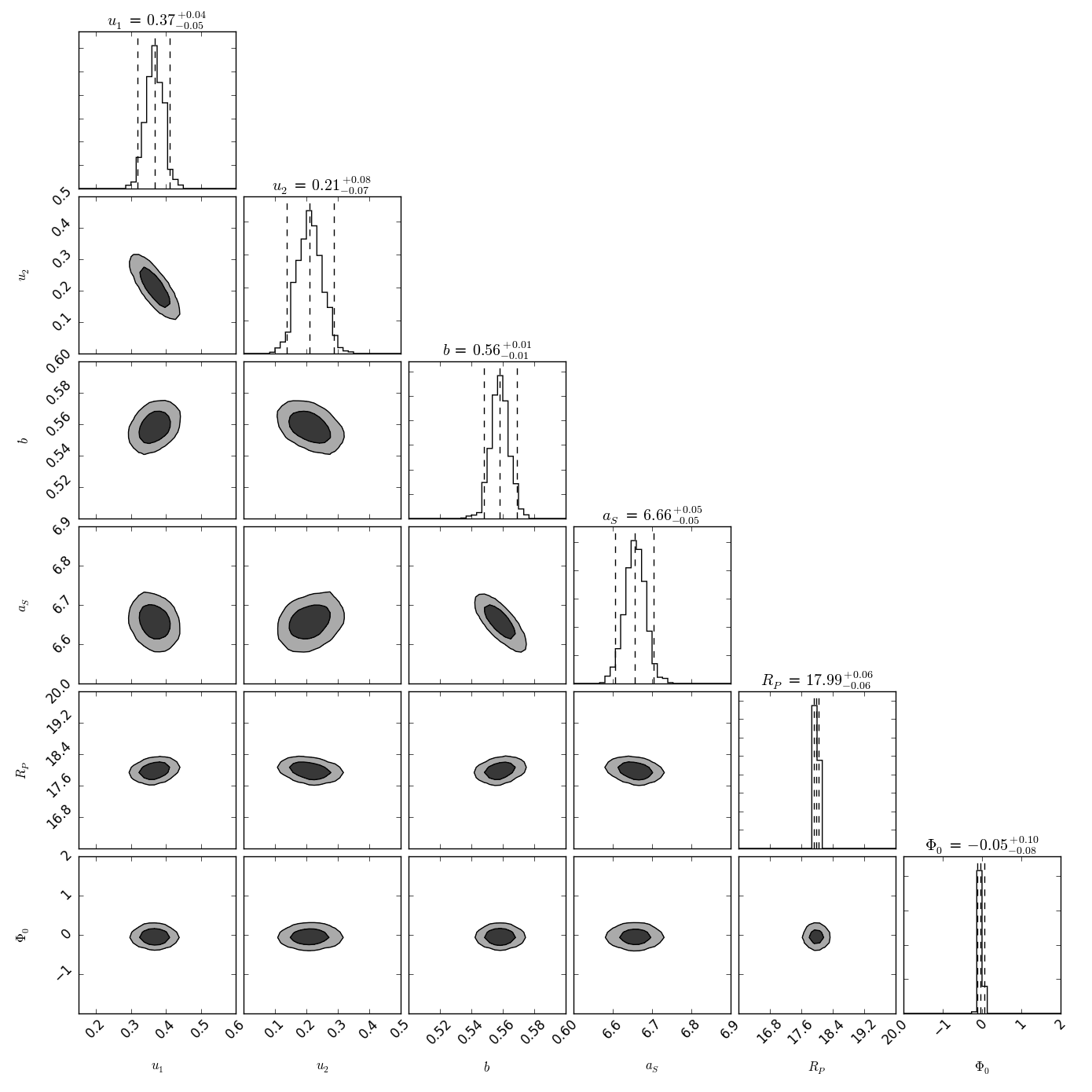}\\
\caption{Kepler-7b posterior projections for the symmetric model with free $\Phi_0$ and $h_{\rm{cloud}}$=0. Dashed vertical lines represent marginalized 95\,\% credibility regions. Smoothed 68\,\% and 95\,\% credibility regions in dark and light grey shade, respectively.}
\label{K7_triangle_t}
\end{center}
\end{figure*}

\begin{figure*}
\begin{center}
\includegraphics[width=450pt]{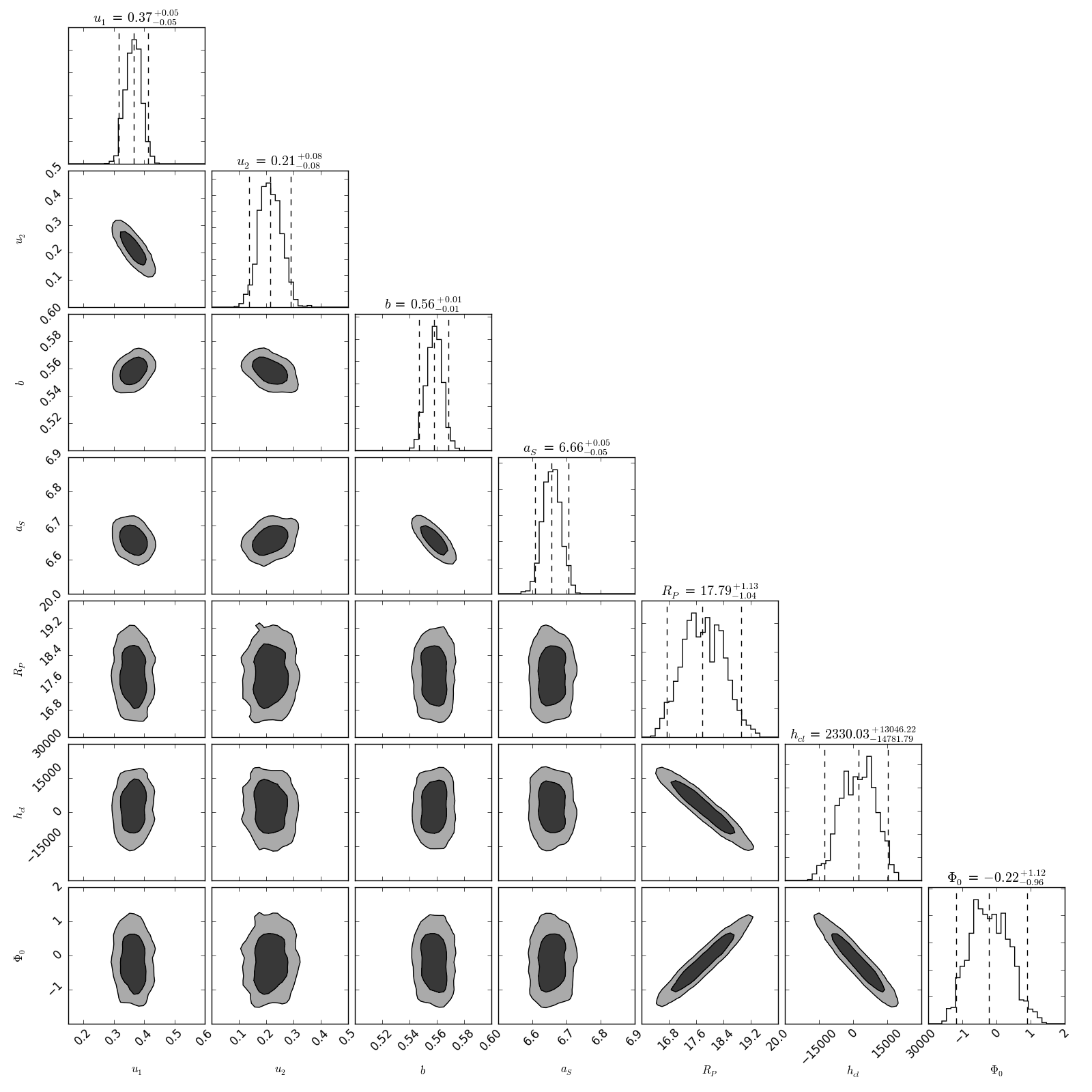}\\
\caption{Kepler-7b posterior projections for the asymmetric model. Dashed vertical lines represent marginalized 95\,\% credibility regions. Smoothed 68\,\% and 95\,\% credibility regions in dark and light grey shade, respectively.}
\label{K7_triangle_asym}
\end{center}
\end{figure*}


%
%
%
%

\begin{figure*}
\begin{center}
\includegraphics[width=450pt]{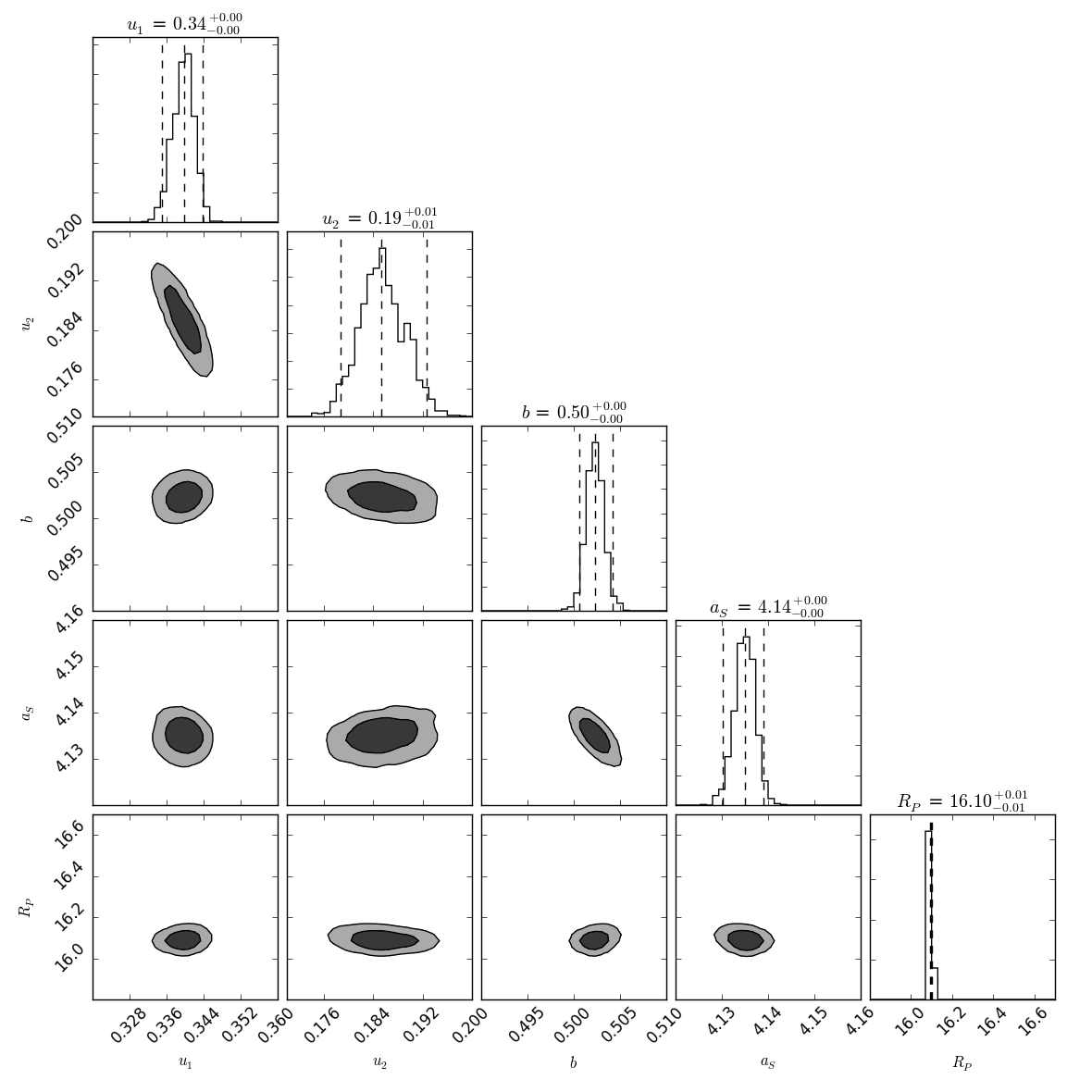}\\
\caption{HAT-P-7b posterior projections for the symmetric model ($\Phi_0$=0, $h_{\rm{cloud}}$=0). Dashed vertical lines represent marginalized 95\,\% credibility regions. Smoothed 68\,\% and 95\,\% credibility regions in dark and light grey shade, respectively.}
\label{H7_triangle_no}
\end{center}
\end{figure*}

\begin{figure*}
\begin{center}
\includegraphics[width=450pt]{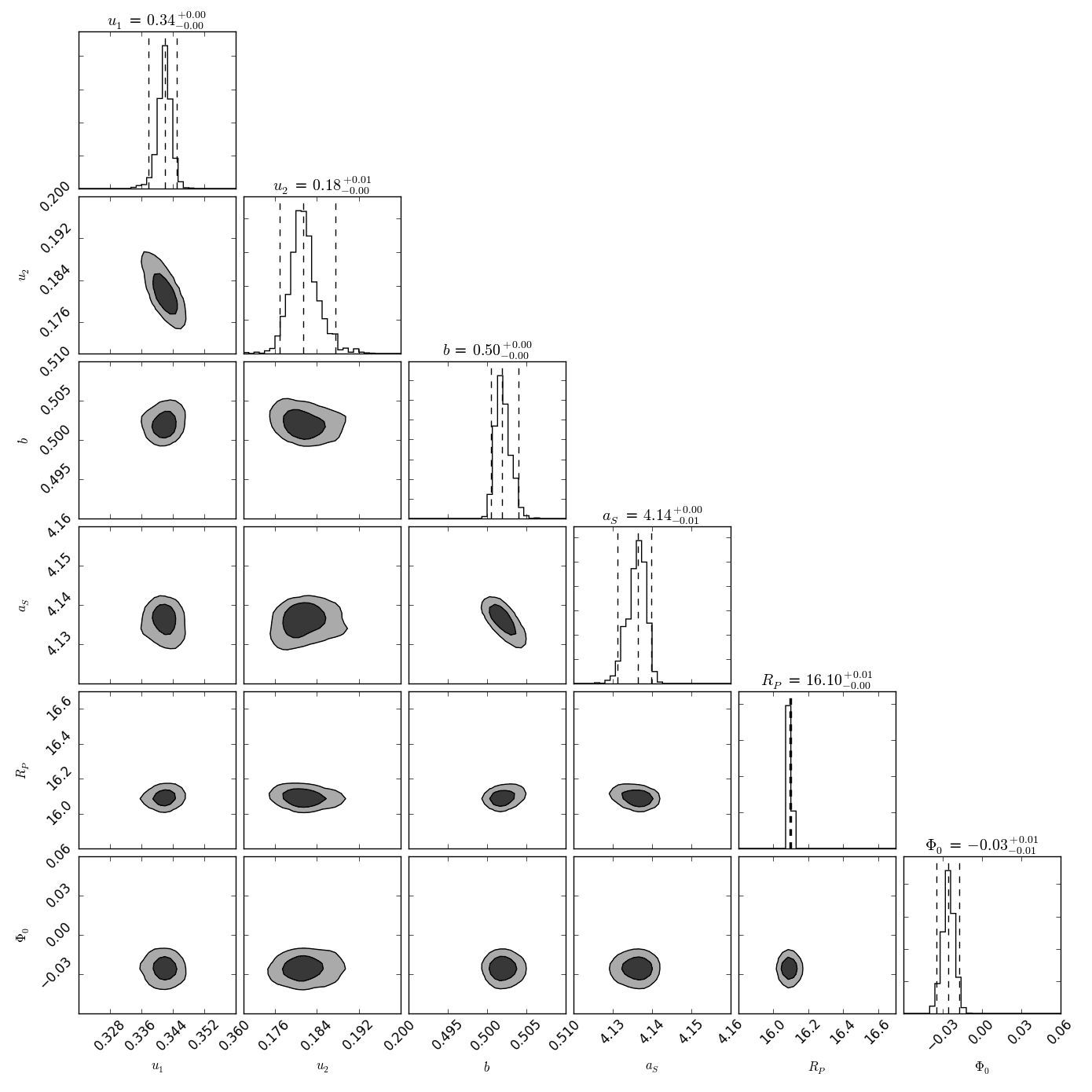}\\
\caption{HAT-P-7b posterior projections for the symmetric model with free $\Phi_0$ and $h_{\rm{cloud}}$=0. Dashed vertical lines represent marginalized 95\,\% credibility regions. Smoothed 68\,\% and 95\,\% credibility regions in dark and light grey shade, respectively.}
\label{H7_triangle_t}
\end{center}
\end{figure*}

\begin{figure*}
\begin{center}
\includegraphics[width=450pt]{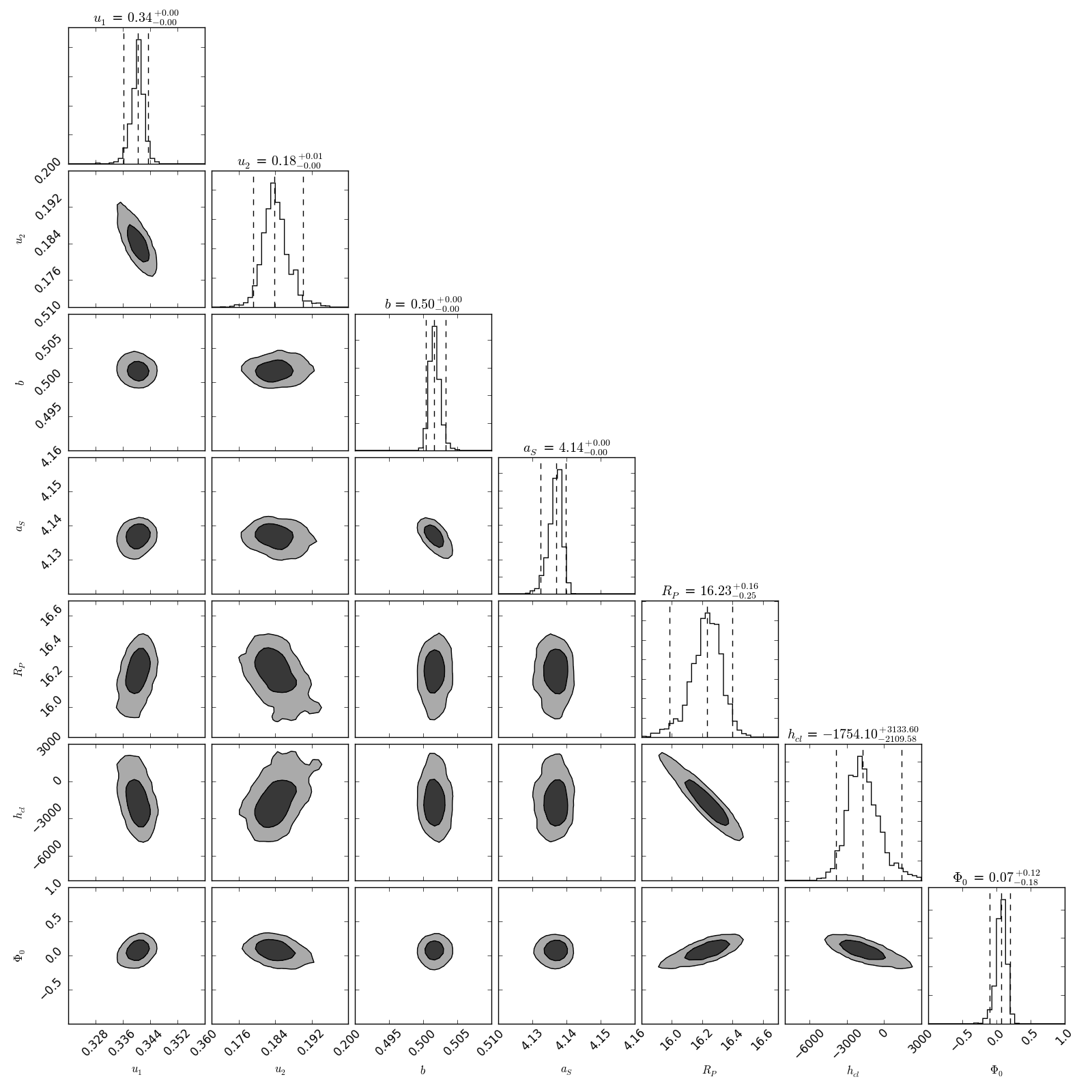}\\
\caption{HAT-P-7b posterior projections for the asymmetric model. Dashed vertical lines represent marginalized 95\,\% credibility regions. Smoothed 68\,\% and 95\,\% credibility regions in dark and light grey shade, respectively.}
\label{H7_triangle_asym}
\end{center}
\end{figure*}

\end{document}